\documentclass[structabstract]{aa}  

\usepackage{txfonts}
\usepackage{natbib}
\usepackage{epsfig}
\usepackage{graphicx}
\usepackage{tabularx}
\usepackage[applemac]{inputenc}
\usepackage{enumerate}
\usepackage{pdflscape}
\begin{document}
   \title{Fossil Groups Origins}

   \subtitle{III. Characterization of the sample and observational properties of fossil systems}

   \author{S. Zarattini \inst{1, 2,3}, R. Barrena \inst{1, 2}, M. Girardi \inst{4,5}, N. Castro-Rodriguez\inst{1, 2}, W. Boschin\inst{6}, J. A. L. Aguerri \inst{1, 2}, J. M\'endez-Abreu\inst{1, 2, 18}, R. S\'anchez-Janssen\inst{7}, C. Catalán-Torrecilla\inst{8}, E. M. Corsini\inst{3,9}, C. del Burgo\inst{10}, E. D'Onghia\inst{11,12}, N. Herrera-Ruiz\inst{13}, J. Iglesias-P\'aramo\inst{14,15},  E. Jimenez Bailon\inst{16}, M. Lozada Muñoz\inst{16}, N. Napolitano\inst{17},  \and J. M. Vilchez\inst{14}}
          

     \institute{Instituto de Astrofísica de Canarias, calle Vía Láctea S/N, E-38205 La Laguna, Tenerife, Spain
     \and Universidad de La Laguna, Dept. Astrofísica, E-38206 La Laguna, Tenerife, Spain
     \and Dipartimento di Fisica e Astronomia "G. Galilei", Università degli Studi di Padova,  vicolo dell'Osservatorio 3, I-35122 Padova, Italy
     \and Dipartimento di Fisica-Sezione Astronomia, Università degli Studi di Trieste, via Tiepolo 11, I-34143 Trieste, Italy
     \and INAF - Osservatorio Astronomico di Trieste, via Tiepolo 11, I-34143 Trieste, Italy
     \and Fundación Galileo Galilei - INAF, Rambla José Ana Fernández Pérez 7, E-38712 Breña Baja, La Palma, Spain
      \and NRC Herzberg Institute of Astrophysics, 5071 West Saanich Road, Victoria, BC, V9E 2E7 Canada
      \and Departamento de Astrofísica y CC. de la Atmósfera, Universidad Complutense de Madrid, E-28040 Madrid, Spain
      \and INAF - Osservatorio Astronomico di Padova, vicolo dell'Osservatorio 5, I-35122 Padova, Italy
      \and Instituto Nacional de Astrofísica, Óptica y Electrónica, Luis Enrique Erro 1, Sta. Ma. Tonantzintla, Puebla, México 
     \and Astronomy Department, University of Wisconsin, 475 Charter St., Madison, WI 53706 USA
     \and Alfred P. Sloan Fellow
     \and Astronomisches Institut der Universität Bochum, Universitätsstr. 150, 44801 Bochum, Germany
     \and Instituto de Astrofísica de Andalucía – C.S.I.C., E-18008 Granada, Spain
     \and Centro Astronómico Hispano Alemán, C/ Jesús Durbán Remón 2-2. 04004 Almería, Spain
     \and Instituto de Astronomía Apdo. 70-264, Cd. Universitaria, México DF 04510 México
     \and INAF - Osservatorio Astronomico di Capodimonte, Salita Moiariello 16, I-80131, Napoli, Italy  
     \and School of Physics and Astronomy, University of St Andrews, North Haugh, St Andrews, KY16 9SS, UK
     \\ \email{stefano@iac.es}}
             
   \date{Accepted 26/02/2014}

 
  \abstract
   {Virialized halos grow by the accretion of smaller ones in the cold dark matter scenario. The rate of accretion depends on the different properties of the host halo. Those halos for which this accretion rate was very fast and efficient resulted in systems dominated by a central galaxy surrounded by smaller galaxies at least two magnitude fainter. These galaxy systems are called fossil systems and they can be the fossil relics of ancient galaxy structures.}
   {We started an extensive observational program to characterize a sample of 34 fossil group candidates spanning a broad range of physical properties.}
   {Deep $r-$band images were obtained with the 2.5-m Isaac Newton Telescope and Nordic Optic Telescope. Optical spectroscopic observations were performed at the 3.5-m Telescopio Nazionale Galileo telescope for $\sim$ 1200 galaxies. This new dataset was completed with Sloan Digital Sky Survey Data Release 7 archival data to obtain robust cluster membership and global properties of each fossil group candidate. For each system, we recomputed the magnitude gaps between the two brightest galaxies ($\Delta m_{12}$) and the first and fourth ranked galaxies ($\Delta m_{14}$) within 0.5 $R_{{\rm 200}}$. We consider fossil systems those with $\Delta m_{12} \ge 2$ mag or $\Delta m_{14} \ge 2.5$ mag within the errors.}
   {We find that 15 candidates turned out to be fossil systems. Their observational properties are in agreement with those of non-fossil systems. Both follow the same correlations, but the fossil systems are always extreme cases.    
   In particular, they host the brightest central galaxies and the fraction of total galaxy light enclosed in the brightest group galaxy is larger in fossil than in non-fossil systems. Finally, we confirm the existence of genuine fossil clusters.}
   {Combining our results with others in the literature, we favor the merging scenario in which fossil systems formed due to mergers of $L^\ast$ galaxies. The large magnitude gap is a consequence of the extreme merger ratio within fossil systems and therefore it is an evolutionary effect. Moreover, we suggest that at least one fossil group candidates in our sample could represent a transitional fossil stage. This system could have been fossil in the past, but not now due to the recent accretion of another group of galaxies.}
   \keywords{Galaxies: groups: general – Galaxies: clusters: general – Galaxies: formation – Galaxies: evolution – Galaxies: distances and redshifts – Galaxies: elliptical and lenticular, cD}
	\authorrunning{Zarattini et al.}
	\titlerunning{Observational properties of fossil groups}
   \maketitle
%


\section{Introduction}
\label{intro}

Fossil systems are group- or cluster-sized \citep{mendes06,cypriano06} objects whose luminosity is dominated by a very massive central galaxy. In the current cold dark matter (CDM) scenario, these objects formed hierarchically at an early epoch of the Universe and then slowly evolved until present day. That is the reason why they are called {\it fossils}. 

The study of this particular kind of objects started two decades ago, when \citet{ponman94} suggested that RX-J1340.6+4018 was probably the remains of an ancient group of galaxies. Later, \citet{jones03} gave the first observational definition of Fossil Groups (FGs) as systems characterized by a magnitude gap larger than 2 mag in the $r-$band between the two brightest galaxies of the system within half the virial radius. Moreover, the central galaxy should be surrounded by a diffuse X-ray halo, with a luminosity of at least $L_{\rm X} > 10^{42}$ $h_{50}^{-2}$ erg ${\rm s^{-1}}$, with the aim of excluding bright isolated galaxies.

Many optical and X-ray observational properties of FGs have been studied, but always on small samples or individual systems. These properties can be grouped in: (i) properties of the intracluster hot component; (ii) properties of the galaxy population; and (iii) properties of the brightest group galaxy (hereafter BGG). Referring to the hot gas component, fossil and non-fossil systems generally show a similar $L_{\rm X}-T_{\rm X}$ relation \citep[see][]{khos07,harrison12}. Differences in scaling relations that combine both optical and X-ray properties were detected. In particular, some authors found different  relations in optical vs X-ray luminosity ($L_{{\rm opt}}-L_{\rm X}$), X-ray luminosity vs velocity dispersion of the clusters galaxies ($L_{\rm X}-\sigma _{{\rm v}}$), and X-ray temperature vs velocity dispersion ($T_{\rm X}-\sigma _{{\rm v}}$). In these works, for any given L$_{{\rm opt}}$, FGs are more luminous and hotter in the X-rays than normal groups or clusters. These differences were interpreted as a deficit formation of $L^\ast$ galaxies in FGs \citep[see][]{proctor11}. In contrast, other authors such as \citet{voev10} and \citet{harrison12} did not find any different relation between X-ray and optical quantities for FGs and normal groups and clusters. They claimed that the previous differences were due to  observational biases in the selection of FGs or inhomogeneity between the FGs and the comparison sample. In addition, high $S/N$ and high resolution X-ray observations of fossil systems seem to confirm that fossil systems are formed inside high centrally concentrated dark matter (DM) halos \citep{sun04,khos06}, with large mass-to-light ratios, which could indicate an early formation. Nevertheless, most of the fossil systems do not show cooling cores \citep[but see also][]{democles10} as normal clusters, suggesting that strong heating mechanisms, such as AGN feedback or cluster mergers, could heat the central regions of their DM halos \citep{sun04,khos04,khos06,mendes09}.  

The galaxy luminosity function (hereafter LF) is a powerful tool for studying the galaxy population in clusters. In the past, several works investigated the galaxy LF in fossil systems. They found that the LF of these objects are well fitted by single Schechter function, but there is a large variety of values in the faint-end slope ($\alpha$) of the LFs of FGs. In particular, the values of $\alpha$ measured goes from $-1.6$ to $-0.6$ \citep[see][]{cypriano06,khos06,mendes06,mendes09,zibetti09,aguerri11,lieder13}.  Unfortunately, all these studies were performed on single FGs or very small samples, and a systematic study of LFs of statistically meaningful samples of FGs remains to be done. 
  
The brightest central galaxies of fossil systems are amongst the most massive and luminous galaxies known in the Universe. In fact, the luminosity and the fraction of light contained in the BGGs  correlate with the magnitude gap \citep{harrison12}.  Some observations  \citep{khos06} show that these objects are different from both isolated elliptical galaxies and central galaxies in non-fossil clusters in the sense that they have disky isophotes in the centre and their luminosity correlates with velocity dispersion, while other authors  \citep{labarbera09, jairo12} found no differences in isophotal shapes between fossil and non fossil central galaxies. In \citet{jairo12} we analised deep K-band images of 20 BGGs in fossil and non-fossil systems and showed that these galaxies follow the tilted fundamental plane of normal ellipticals \citep[see ][]{bernardi11}. This fact suggests that BGGs are dynamically relaxed systems that suffer dissipational mergers during their formation. On the other hand, they depart from both Faber-Jackson and luminosity-size relations. In particular, BGGs have larger effective radii and smaller velocity dispersions than those predicted by these relations. We infer that BGGs grew throughout dissipational mergers in an early stage of their evolution and then assembled the bulk of their mass through subsequent dry mergers. Nevertheless, stellar population studies of BGGs in fossil systems suggest that their age, metallicity and $\alpha$-enhancement are similar to those of bright ellipticals field galaxies \citep[see][]{labarbera09,paul13}.

In numerical simulations, FGs are found to be a particular case of structure formation. They are supposed to be located in highly concentrated DM halos, so that they can assembly half of their DM mass at {\it z} $>$ 1. Then, the FGs grow via minor mergers only, and only accrete $\approx$ 1 galaxy from {\it z} $\approx 1$ down to present time, while regular groups accrete about three galaxies in the same range of time \citep{vonbenda08}.  \citet{dariush07} show that the mass assembled at any redshift is higher in fossil than in non-fossil systems. This means that the formation time is, on average, shorter for FGs than for regular systems \citep{donghia05, vonbenda08}, leaving to FGs enough time to merge $L^\ast$ galaxies in one very massive central object. In fact, simulations predict that the timescale for merging via dynamical friction is inversely proportional to the mass of the galaxy, thus favouring the merging of larger objects. So, the dynamical friction would be responsible for the lack of $L^\ast$ galaxies which is reflected in the requested magnitude gap of the observational definition. Moreover, to enhance the high efficiency in the merging process, FGs should have particular dynamical properties, such as the location of massive satellites on orbits with low angular momentum \citep[see][]{sommer06}. So, a combination of high mass satellite and low angular momentum orbits boosts the efficiency of the merging process \citep{boylan08}. Recently, \citet{lidman13} demonstrate that the growth of the BCGs since z$\sim$1 is mainly due to major mergers, suggesting that this could be the dominant mechanism in accreting the mass of central galaxies in cluster and thus supporting indirectly the merging scenario for fossil systems, which would differ from regular clusters only for the early time formation. Nevertheless, this evolutionary picture in which fossil groups became fossils in the early Universe and then evolved undisturbed is not the only proposed scenario. In the framework of the merging scenario \citet{diaz08} suggest that first ranked galaxies in fossil systems has the last major merger later than non-fossil ones. This means that the formation of large magnitude gaps as those in nowadays fossil systems is a long term process. In addition, \citet{vonbenda08} predict that the fossilness could be a transitional status of some systems. Thus, some fossil systems have become non-fossil ones in recent epoch due to accretion of nearby galaxy groups.

An alternative formation scenario in which the magnitude gap of the systems appears at the beginning of the formation process can also explain the reported observational properties. This is the so called monolithic scenario, in which fossil systems formed with a top heavy LF. In this scenario, the magnitude gap is due to a primordial deficient formation of $L^\ast$ galaxies \citep{mul99}.

All the observational results presented in the literature are limited by the small number of FGs known in the literature. A more general study of fossil systems is needed in order to discriminate between these two formation scenarios. For this reason, we started an extensive observational program called Fossil Groups Origins (FOGO), aimed at carrying out a large, systematic and multiwavelenght study of a sample of 34 FGs candidates identified by \citet{santos07}. The specific goals of the program include mass and dynamics of FGs, properties of their galaxy populations, formation of the central galaxies and their connection with the intragroup medium, properties of the extended diffuse light, and agreement with old and new simulations. The details of the project are resumed in the first paper of the series \citep[][]{aguerri11}. The structural properties of the BGGs in fossil and non-fossil systems were shown in the second paper \citep[][]{jairo12}. The L$_{\rm X}$-L$_{{\rm opt}}$ relation of fossil and normal systems will be presented in a forthcoming paper (Girardi et al., {\it in prep}). This is the third paper of the series, devoted to the characterization of the sample. In particular, we recomputed the magnitude gaps of the systems by using new spectroscopic redshift measurements. These new data provide us robust cluster membership and global properties for the systems. Only 15 out of 34 turned out to be fossil systems according with the standard definition \citep[see][]{jones03}. We have also analised the relations between central galaxies in FGs and non-FGs and other quantities such as magnitude gaps and velocity dispersion of the host halo. FGs follow the same relations than non-FGs, but they are extreme cases.

The paper is organized as follows. The description of the sample is given in Section 2. The available dataset is shown in Section 3. Radial velocities determination are presented in Section 4. The results are given in Section 5. Sections 6 and 7 report the discussion and conclusions of the paper, respectively. Unless otherwise stated, we give errors at the 68\% confidence level. Throughout this paper, we use $H_0=70$ km s$^{-1}$ Mpc$^{-1}$, $\Omega_{\Lambda}=0.7$ and $\Omega_M=0.3$


\section{Description of the sample}
\label{sample}
\begin{figure*}
\centering
\includegraphics[bb=20 140 600 305]{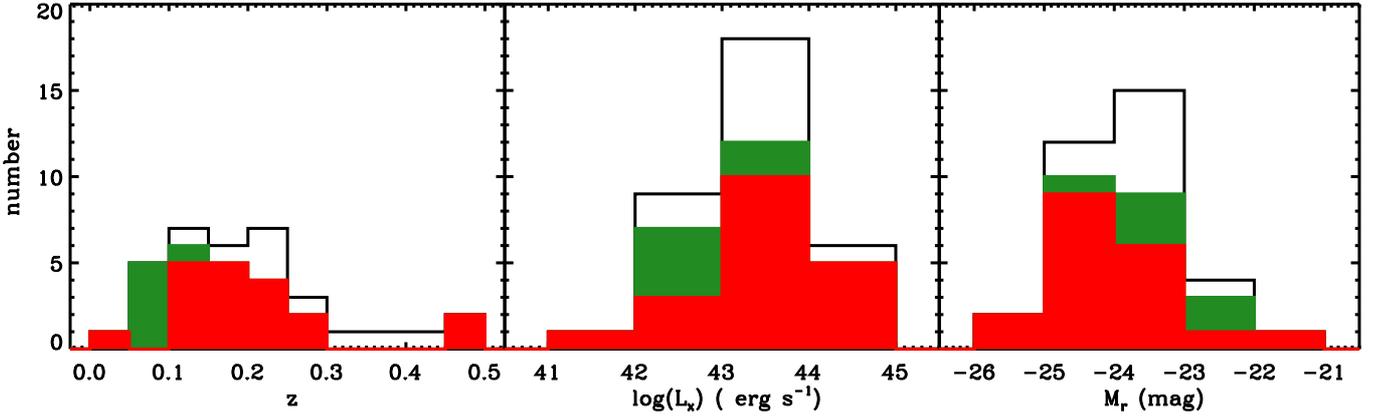}
\caption{Distribution of redshifts (left panel),  X-ray luminosities (central panel), and absolute magnitudes of the BGGs (right panel) for the sample of 34 FG candidates by \citet{santos07} (black line) and for our subsample of 25 FG candidates either with our own (red line) or with SDSS spectroscopy (green line).}
\label{fig1}
\end{figure*}

The FOGO sample is based on the \citet{santos07} FG candidates selected from the Sloan Digital Sky Survey Data Release 5 \citep[SDSS DR5,][]{DR5}. 
\citet{santos07} selected 112.510 galaxies brighter than $r=19$ in the Luminous Red Galaxy catalog \citep[LRG,][]{LRG}. A cross-match with the Rosat All Sky Survey catalog \citep[RASS,][]{RASS} was performed to look for the presence of a diffuse X-ray halo of at least $10^{42}$ erg s$^{-1}$ and closer than 0$\farcm$5 from the position of each LRG. FInally, they looked for the brightest companions of each of the remaining LRGs within a fixed radius of $0.5h_{70}^{-1}$ Mpc to satisfy the magnitude gap $\Delta m_{12} \ge 2$ between the two brightest galaxies of the group. The final catalog comprises 34 FG candidates with some unique characteristics: the sample spans the last 5 Gyr of galaxy evolution ($0\le z\le 0.5$), it has a wide range of X-ray luminosities and therefore masses, and the absolute magnitude of the central BGG covers a large range ($-25.3\le M_{r}\le -21.3$).

In this work we present the analysis of the 34 systems of the sample of \citet{santos07}. For each system we were able to compute new $\Delta m_{12}$ and $\Delta m_{14}$ gaps combining our deep $r-$band images with photometric data from the SDSS DR7. We measure the LOS velocity dispersion for those systems with at least 10 members within $R_{200}$ \footnote{The radius $R_{{\rm \delta}}$ is the radius of a sphere with mass over density $\delta$ times the critical density at the redshift of the galaxy system}. This subsample is formed by 24 groups with redshift obtained mainly from our own spectroscopy ($\sim 1200$ new velocities, see Sect. \ref{obs}). In Fig. \ref{fig1} we show the distribution of redshifts, X-ray luminosities and absolute magnitude of the 34 BGGs of the sample by \citet{santos07} and of our subsample of 24 objects plus FGS28, for which only one member galaxy is found within $R_{200}$. The Kolmogorov-Smirnov test confirms that our subsample of 25 and the whole sample of 34 FG candidates by \citet{santos07} come from the same parent distribution.

Hereafter, we identify each system using the notation FGS + ID, where ID is the id number in Table 1 of \citet{santos07}.


\section{The data}
\label{data}

\subsection{Photometric data}
\label{opt}

Deep images for 22 of the FG candidates were obtained using the 2.5-m Nordic Optical Telescope (NOT) at the Roque de los Muchachos Observatory (ORM, La Palma, Spain) in the period between 2008-2011. We used Andalucia Faint Object Spectrograph and Camera (ALFOSC) in imaging mode with SDSS $r-$band filter. The detector was a CCD of 2048$\times$2048 pixels with a plate scale of 0\farcs19 pixel$^{-1}$. For other 10 candidates, images in the same band were taken at the 2.5-m Isaac Newton Telescope (INT) at the ORM in the same period using the Wide Field Camera (WFC). It consists of 4 2000$\times$4000 CCDs with a scale of 0\farcs33 pixel$^{-1}$. All the images were obtained under photometric conditions, and the mean value of the seeing FWHM was 1\farcs0. Only FGS27 and FGS33 were observed under bad seeing conditions. Their final combined images have a seeing FWHM $\ge$ 2\farcs and therefore they were replaced with SDSS images. For FGS09 and FGS26 it was impossible to obtain deep images due to the presence of a very bright star located close to their BGGs. SDSS photometric images were also used for these two systems.

Data reduction was performed using standard IRAF\footnote{IRAF is distributed by the National Optical Astronomy Observatories, which are operated by the Association of Universities for Research in Astronomy, Inc., under cooperative agreement with the National Science Foundation.} routines, correcting for bias and flat field. In most of the cases, after these corrections, we detected some residual light. In order to achieve the best possible flat field correction, we obtained a super-flat field using a combination of the scientific images. Then we corrected once again the scientific images with such a new super-flat field \citep[see ][for details]{aguerri11}. The images were combined and calibrated using SDSS data of the not saturated stars available in the field of view. The typical RMS error of the calibration is 0.08 mag.

\subsection{Spectroscopic data}
\label{obs}

We used the SDSS DR6 \citep{DR6} photometric redshifts ($z_{{\rm phot}}$) to select reliable targets for multi-object spectroscopy (MOS). For each FG candidate we downloaded a catalog with all galaxies brighter than $m_r = 22$ within a radius of 30 arcmin around each BGG. The $m_r$ value represents the completeness limit of the photometric catalog of the SDSS, the selected radius is larger than a virial radius for all our FG candidates. Then, we considered as possible targets only galaxies with photometric redshift within the range of $\Delta z_{{\rm phot}} \pm 0.15$ from the spectroscopic redshift of the BGG. This value was chosen because the typical error on photometric redshift in the SDSS DR6 is about 0.1. Finally, we visually selected the targets trying to maximize the number of slits per mask. We also put 60 slits on galaxies with spectroscopic redshift in the SDSS DR6 for a comparison. We observed a total of 51 masks with on average 30 slits per mask. We obtained 1227 spectra with a $S/N \ge 5$, which is enough to measure the line-of-sight (LOS) velocity of the galaxies. 

MOS observations were performed under {\it International Time Program} (ITP) of the ORM in the period between 2008-2010. Additional observations were done under one Italian and two Spanish {\it Time Allocation Comitee} (TAC) runs between 2008-2012. The data were taken at the 3.5-m Telescopio Nazionale Galileo (TNG) telescope using Device Optimized for the LOw RESolution (DOLORES) in the MOS mode. The instrument has a CCD of $2048\times2048$ pixels with  a pixel size of 13.5 $\mu$m and a $0\farcs252$ pixel$^{-1}$ scale. We used the LR-B Grism with a dispersion of 187 $\AA$ mm$^{-1}$, together with $1\farcs6$ slits. This led to a final resolution of $R=365$ in the wavelength range 3000-8430 \AA. 
The typical exposure time was of 3$\times$1800s per mask and the mean FWHM of the seeing was $1\farcs0$.  

We performed the data reduction using standard IRAF procedures. We did not correct for bias and flat field because the uncorrected spectra result less noisy than the corrected ones. In particular, the measured LOS velocities are the same in both the corrected and uncorrected spectra, but the uncertainties are larger when the bias and flat field corrections are applied. The cosmic rays correction was performed during the combining process of the different exposures we obtained for each mask. The sky was evaluated by measuring the median value in the outer regions of each spectrum. To perform the wavelength calibration we used two different lamps (Ne+He and Ne+Hg) to have arc lines in both the red and blue part of the spectrum.  The typical uncertainty of the wavelength calibration was $< 0.1$ $\AA$ (RMS).
Finally, we corrected for the instrumental distortions by measuring the [OI]$\lambda$5577 $\AA$ sky line. This is crucial since we divided each mask exposure in individual exposures of 1800s, sometimes taken in different days or runs. The mean error associated to the instrumental distortions is 0.85 $\AA$ (which corresponds to 45 km s$^{-1}$), but it can be as large as 8 $\AA$ (450 km s$^{-1}$). We corrected all the measured LOS velocities to take into account for the systematic error due to the instrumental distortions.


\section{Redshift catalog}
\label{redshift}

\subsection{Line of sight velocity measurement}
\label{int_redshift}
\begin{figure*}
\centering
\includegraphics{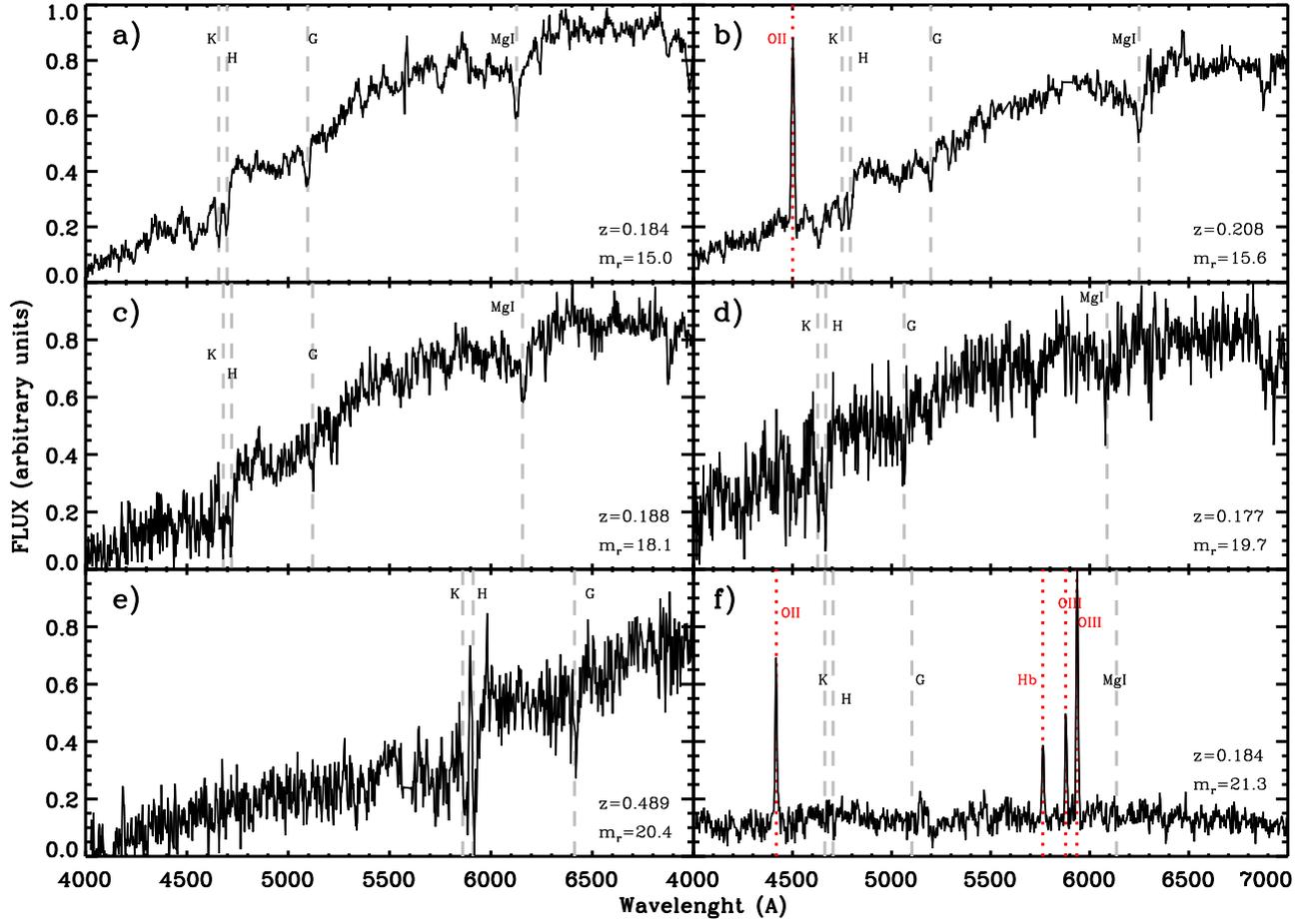}
\caption{Examples of spectra with decreasing S/N (from panel a to panel f) from our dataset. The main absorption and emission lines are marked with dashed grey and dotted red lines, respectively.}.
\label{spectra}
\end{figure*}

The LOS velocities were measured using the cross-correlation technique \citep{ton79} implemented in the IRAF task XCSAO of the package RVSAO\footnote{RVSAO was developed at the Smithsonian Astrophysical Observatory Telescope Data Center.}. For each spectrum the task performs a cross-correlation with six templates \citep{ken92}, corresponding to different types of galaxies (E, S0, Sa, Sb, Sc, Irr). The template with the highest value of the S/N of the cross-correlation peak was chosen. In addition, we visually inspected all the spectra to verify the velocity determination. In most of the cases the LOS velocity was obtained from the absorption lines. Nevertheless, in some spectra with low $S/N$ (especially for faint objects with $m_r > 20.5$) the emission lines were measured with the IRAF task EMSAO to obtain the LOS velocity. In Fig. \ref{spectra} the absorption lines are detectable in the five brightest objects but not in the faintest one. The latter is actually the only galaxy with $m_r > 21$ for which we measured the LOS velocity.
The nominal uncertainties given by the cross--correlation algorithm are known to be
smaller than the true ones \citep[see, ][]{mal92,bar94,ell94,qui00}.  The uncertainties obtained through the RVSAO procedure were multiplied by a
factor 2, following previous analyses \citep[][ and references therein]{bar09} on data acquired with the
same instrumental setup and with comparable quality or our own. Moreover, to be conservative, we assumed the
largest between the formal uncertainty and 100 km s$^{-1}$ for the LOS
velocities measured with the EMSAO procedure .  We adopted the weighted mean of the different determinations and the corresponding error for the galaxies with
repeated measurements . The RMS of this difference for 48 galaxies is 107 km s$^{-1}$.

\subsection{Additional line of sight velocities}
\label{errors}

\begin{figure}
\centering
\includegraphics[width=0.5\textwidth]{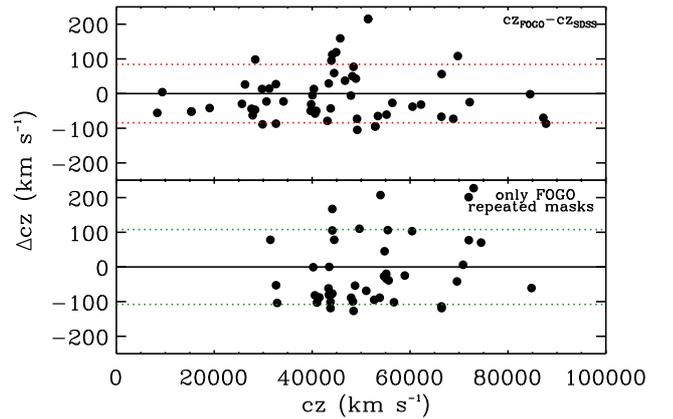}
\caption{Differences in LOS velocity for galaxies with both FOGO and SDSS measurements (top panel) and for galaxies with repeated FOGO measurements (bottom panel). Dotted lines represent the $1\sigma$ scatter of the data.}
\label{errori}
\end{figure}

In order to have the largest possible number of LOS velocities for the 34 FG candidates, we added all available redshifts within $R_{200}$ from the SDSS-DR7. Figure \ref{errori} shows the comparison between our and SDSS LOS velocity measures for the 60 galaxies for which both values are available. The RMS of the difference between the two values is 84 km s$^{-1}$, which is consistent with the results of Sect. \ref{int_redshift}. Finally, for FGS05 we added the LOS velocities given by \citet{A697} and obtained with the same instrumental setup and data analysis.

\subsection{Spectroscopic completeness}
\label{compl}
\begin{figure}
\centering
\includegraphics[width=0.5\textwidth]{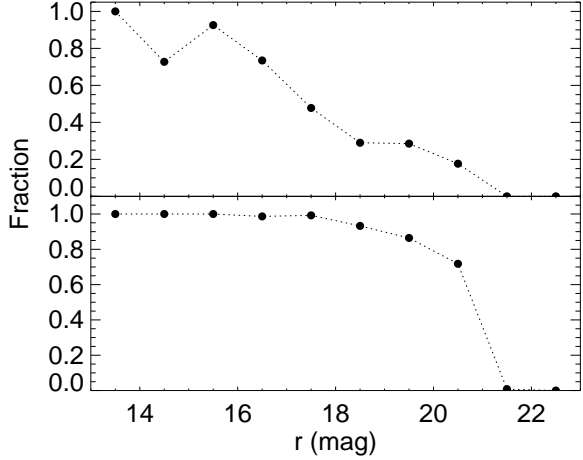}
\caption{Spectroscopic completeness of the 25 FG candidates with either FOGO or SDSS spectroscopy (top panel) and success rate of the 18 FG candidates with FOGO spectroscopy only (bottom panel) as a function of $r-$band magnitude.}
\label{completeness}
\end{figure}

The completeness of the spectroscopic sample is a crucial parameter since it is used in the derivation of several quantities, such as the spectroscopic luminosity function. For each magnitude bin we defined our completeness as the ratio between the number of galaxies of the 25 FG candidates for which we were able to obtain a redshift ($N_z$) from either the FOGO or SDSS spectroscopy and the number of targets ($N_{z_{{\rm phot}}}$, see Sec. \ref{obs}):

\begin{equation}
C=\frac{N_z}{N_{z_{\rm phot}}}.
\end{equation}
In Fig. \ref{completeness} we show our completeness as a function of the $r-$band magnitude. Our sample is more than 70\% complete down to $m_r=17$ and more than 50\% complete at $m_r=18$.

In a similar way, for each magnitude bin, we defined the success rate as the ratio between the number of galaxies of the 18 FG candidates for which we are able to measure a redshift with our own spectroscopy ($N_{z_{{\rm our}}}$) and the total number of observed galaxies ($N_{{\rm obs}}$):

\begin{equation}
SR=\frac{ N_{z_{\rm our}} } { N_{{\rm obs}} }.
\end{equation}
Figure \ref{completeness} also shows the success rate as function of the $r-$band magnitude. Notice that we have a success rate larger than 75\% for objects with $m_r < 21$. The success rate decreases abruptly $m_r > 21$. Thus, we conclude that the adopted combination of instrumental setup and exposure time is effective for measuring the redshift of galaxies with $m_r \leq 21$.

\section{Results}
\label{results}

\subsection{System membership}
\label{memb}

\begin{figure*}
\centering
\includegraphics[width=1.0\textwidth]{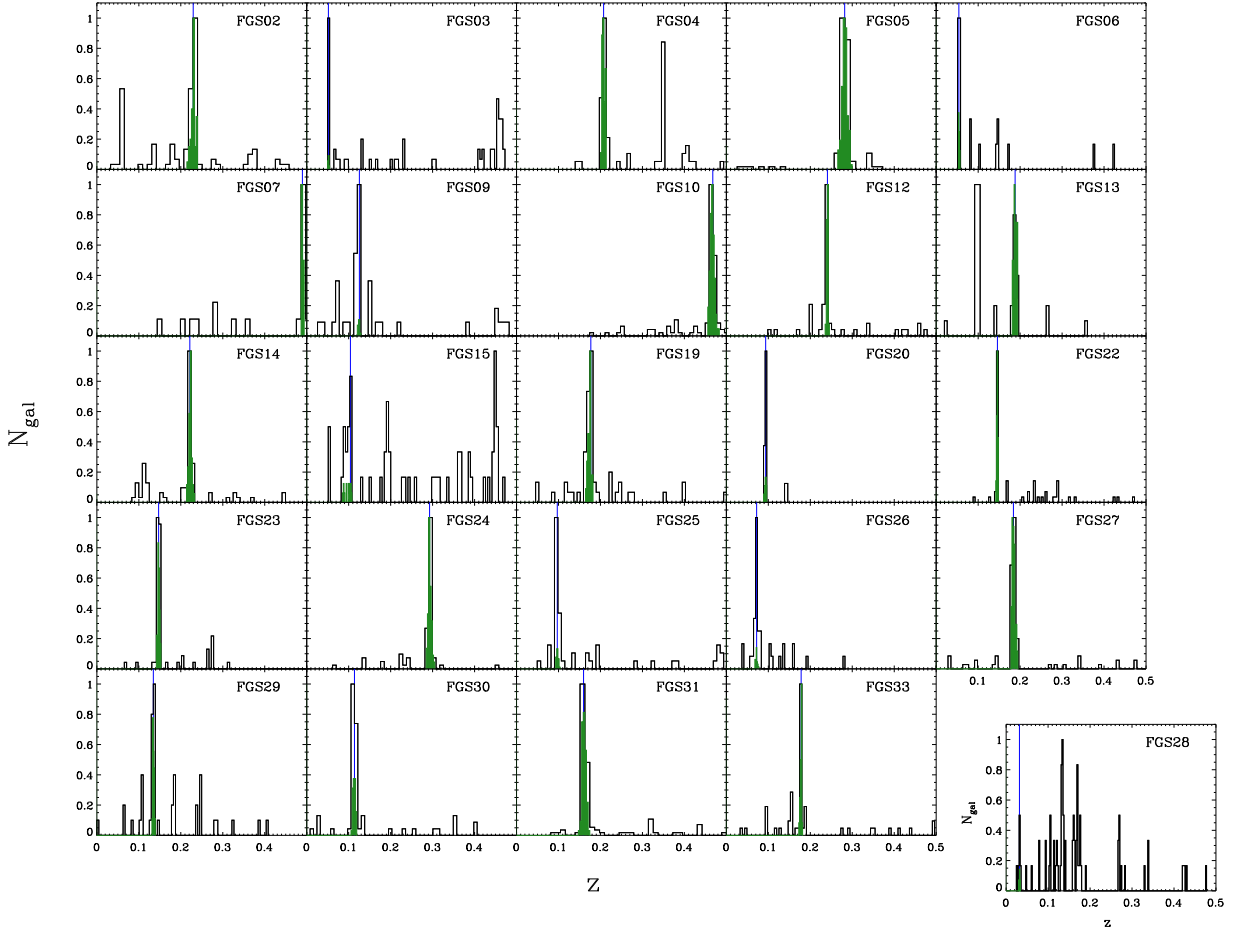}
\caption{Identification of the members of the 25 FG candidates in the redshift space. For each system we show all the galaxies in the field of view (black histogram) as well as the galaxies identified as members using the DEDICA procedure (green histogram) associated to the BGG (blue line). The peculiar case of FGS28 is shown apart.}
\label{vel_hist}
\end{figure*}

\begin{figure*}
\centering
\includegraphics[width=1.0\textwidth]{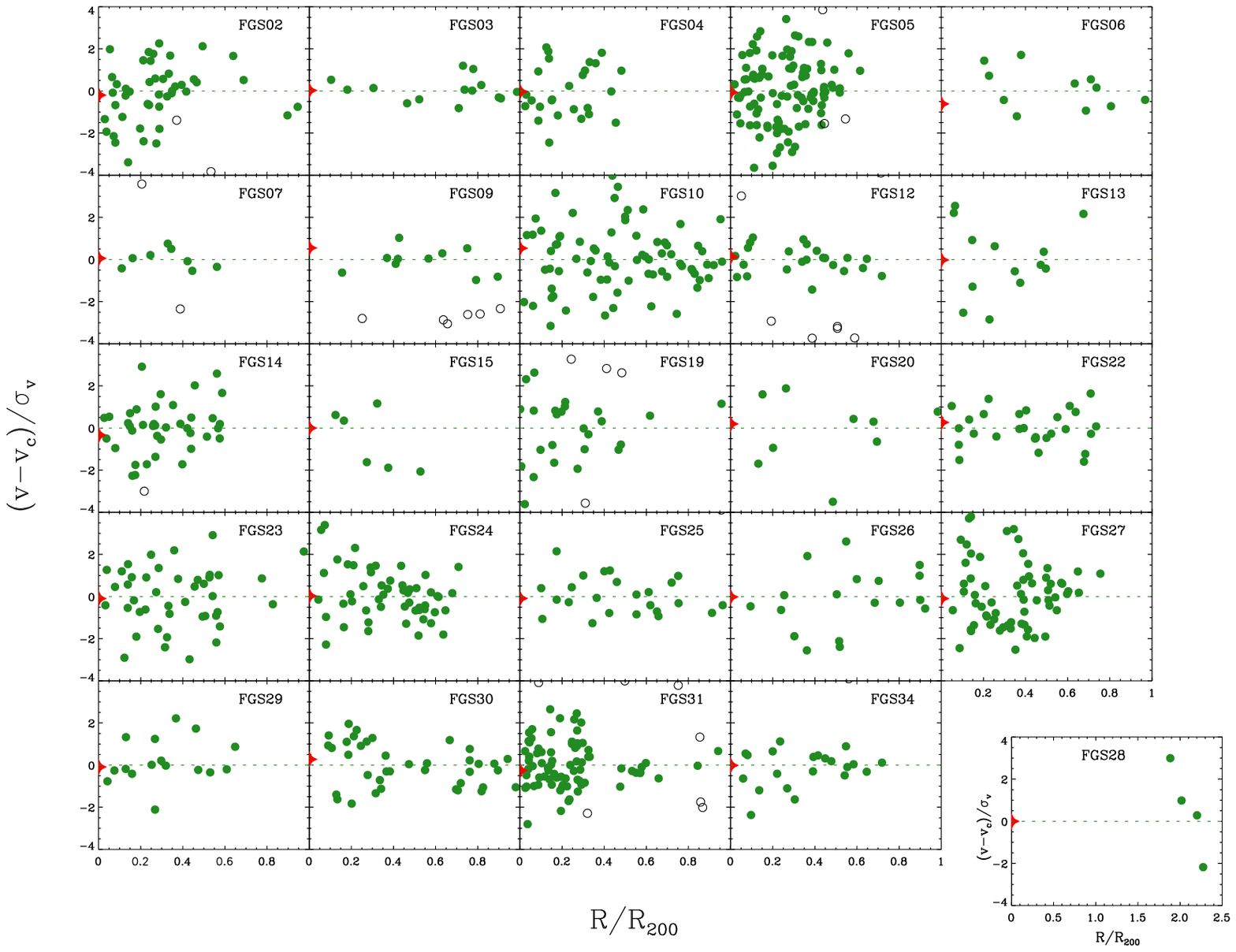}
\caption{Velocity-position diagram of the sample of FG candidates. Green filled and black open circles mark the member and non-member galaxies, respectively. The red star indicates the BGG. The peculiar case of FGS28 is shown apart.}
\label{vel_pos}
\end{figure*}

The identification of systems and the membership of individual
galaxies were obtained using a two-step procedure applied to the
galaxies in the region within $R_{{\rm 200}}$. First, we used DEDICA \citep[][1993]{dedica1}, which is an
adaptive kernel procedure that works under the assumption that a cluster corresponds to a local maximum in the
density of galaxies. Then, we adopted the likelihood ratio test \citep{Materne} to assign a membership probability to
each single galaxy relatively to an identified cluster.

According to the DEDICA procedure, each FG candidate was detected as a very significant peak (at a
confidence level $>99\%$) at the redshift corresponding to that of the BGG. Only FGS14, FGS23, and FGS26 were detected as two close peaks (with $\Delta v$ $<$ 1500 km s$^{-1}$ in the rest frame). For each FG candidate, the redshift of the BGG, the redshift distribution of the galaxies, and the redshift peak associated to the BGG are shown in Fig. \ref{vel_hist}. Some structures are isolated (e.g., FGS20) while others present clear foreground (e.g, FGS07) or background
(e.g., FGS15) contamination. The
corresponding members were then identified using the distance-velocity
diagram (Fig. \ref{vel_pos}), which consists in the so called "shifting gapper"
method \citep{fadda96,girardi96}.  This procedure
rejects galaxies within a fixed distance bin that are too far in velocity from the main body of
the system. The distance bin is shifted outwards out to $R_{200}$. The procedure was then iterated until the number of 
cluster members converged to a stable value. Following \citet{fadda96}, we used a velocity gap of 1000 km s$^{-1}$ in the cluster
rest--frame and a distance bin of 0.6 Mpc (or large enough to include 10
galaxies).  In the case of FGS02 we slightly modified the gap value (1100 km s$^{-1}$) to be more conservative and avoid the rejection of a few galaxies at the edge of the system.
For all the systems our own spectroscopic data extend to at least 0.5 $R_{{\rm 200}}$, except for FGS28, FGS30, and FGS31 for which we covered 0.4 $R_{{\rm 200}}$. The membership efficiency, defined as the fraction of confirmed members with respect to the observed targets, turns out to be 59\%.

FGS15 seems to be a peculiar case within the subsample of 25 FG candidates. It has only 13 members spanning a large range in velocity between one another (up to 6000 km s$^{-1}$). Thus, it is not clear if either this system is very massive or it is part of a larger structure, such as a filament. 
Another system with peculiar properties is FGS28. It is the smallest group of the \citet{santos07} sample. It has the faintest BGG and the lowest X-ray luminosity, and it seems much more an isolated giant elliptical galaxy than a group of galaxies. The peak associated to the BGG in the velocity histogram (Fig. \ref{vel_hist}) is not significant and we found only four (possible) members which are at a distance about $2\,R_{200}$ of the group. We argue that this BGG is actually a large and isolated galaxy which is embedded in a high density environment due to the presence of another cluster in the same region. 
For both FGS28 and FGS15, we used the membership only for calculating the magnitude gaps, but not for estimating the LOS velocity dispersion and mass.

\subsection{Cluster global properties}
\label{glob}

In Table \ref{globprop} we present the general properties for each system of the \citet{santos07} sample. We estimated $R_{{\rm 200}}$ from $L_{{\rm X}}$, which is available for each of the 34 FG candidates from the RASS Catalogs. We decided to recalculate $L_{\rm X}$ because of the discrepancies between the values reported by \citet{santos07} and other measurements available in the literature for some well-studied clusters of their sample \citep[e.g., A267 and A697, see ][after the adequate cosmology and band conversions]{Bor00}
For each FG candidate, we took into account the counts from the RASS Bright Source Catalog \citep[RASS-BSC, ][]{vog99} or, alternatively, from the , RASS Faint Source Catalog \citep[RASS-FSC, ][]{vog00} in the broad band 0.1-2.4 keV. We used the total Galactic column density ($N_{\rm H}$) as taken from the NASA's HEASARC $N_{\rm H}$ tool\footnote{http://heasarc.gsfc.nasa.gov/cgi-bin/Tools/w3nh/w3nh.pl} and the redshift $z$ as listed by \citet{santos07}. We used an iterative procedure based on PIMMS\footnote{ftp:${\rm //legacy.gsfc.nasa.gov/software/tools/pimms4\_3.tar.gz}$} \citep{muk93}. Details of the procedure are available in \citet{girardi14}.

With our new $L{\rm _{x}}$ estimates, we were able to calculate $R_{{\rm 500}}$ using the relation proposed by \citet[][see their equation 2 for details]{Bor07}:

\begin{equation}
R_{500}=0.753\, {\rm Mpc}\, h^{-0.544}_{100} \,E(z)^{-1}\, L_{X,44}^{0.228}
\end{equation}
where $E(z)=h(z)/h_0$ and $L_{X,44}$ is the X-ray luminosity in units of $h_{70}^{-2}10^{44}$ erg s${-1}$ in the 0.1-0.24 keV band. 

We calculated $R_{200}=1.516 \times R_{500}$ as prescripted by \citet{Arnaud05}.

We computed the mean LOS velocity dispersion $\sigma _{{\rm v}}$ by using the bi-weight estimator of the ROSTAT package \citep{beers90} for systems with more than 10 members. For the remaining systems we computed $\sigma _{{\rm v}}$ using the bi-weight estimator and shifting gapper. We applied both the cosmological correction and standard correction for velocity uncertainties \citep{danese80}.
By assuming sphericity, dynamical equilibrium and that galaxy distribution traces mass distribution, we followed \citet{girardi98} and \citet{Girardi_rad} to compute the virial mass as:

\begin{equation}
M_{\rm v} = \frac{3\pi}{2G} \sigma_v \,R_{{\rm pv}}-{\rm SPT}
\label{eq:mass}
\end{equation}

where SPT is the surface pressure term correction \citep{pressure}, while $R_{{\rm pv}}$ is two times the projected mean harmonic radius. 
We could not compute $R_{{\rm pv}}$ by using data of observed galaxies since our $z$ data do not cover the whole $R_{{\rm 200}}$ region. Therefore, we used an alternative estimate which is valid for a typical galaxy distribution in clusters \citep[see eq. 13 of ][]{girardi98}.  We assumed 20\% for the SPT correction, as obtained by combining data on many clusters and valid at a radius around $R_{{\rm 200}}$ \citep{carlberg97,girardi98}.

\begin{table*}
\begin{center}
\caption{Global properties of our sample.}
\label{globprop}
\begin{tabularx}{\textwidth}{lcccrrccccc}
\tiny
\\
\hline
\hline
name& ra (J2000)& dec (J2000)& $z$ &$\Delta m_{12}$ & $\Delta m_{14}$& $R_{{\rm 200}}$ & $N_{{\rm vel}}$ & $N_{{\rm memb}}$   & $\sigma_{{\rm v}}$ & $M_{{\rm v}}$   \\ 
          &   (hh:mm:ss) &   (º:':")       &              & (mag)	& (mag)                 &   (Mpc)     &  (\# Gal)  & (\# Gal) &    (km s$^{-1}$) &  (M$_{\odot}$) \\
      \multicolumn{1}{c}{(1)} & (2) & (3) & (4) & (5) & (6) & (7) & (8) & (9) & (10) & (11) \\
            \hline	
FGS01  	& 01:50:21.3	&	-10:05:30.5 	& 0.365     & $>$1.41$\pm$ 0.23        & $>$1.61 $\pm$ 0.19    	&  1.71 & -			&-	& - 		& - \\ 
FGS02*  	& 01:52:42.0	& 	+01:00:25.6 	& 0.230     & $>$2.12$\pm$ 0.33 	  & $>$2.28 $\pm$ 0.33    	& 1.85  &111(65) 	& 42	 &  1263	& 1.87E+15\\ 
FGS03* 	&07:52:44.2	& 	+45:56:57.4 	&  0.052    & 2.09 $\pm$ 0.06 		  & 2.55 $\pm$ 0.08 		&  0.96 &89 (0) 	& 16	 &    259 	& 4.20E+13\\ 
FGS04  	& 08:07:30.8	&	+34:00:41.6	& 0.208     & $>$1.65 $\pm$ 0.27 	  & $>$2.04 $\pm$ 0.25   	& 1.44 &  65 (64) 	& 28	 & 852 	& 6.67E+14\\
FGS05  	& 08:42:57.6	&	+36:21:59.3	& 0.282     & 1.12 $\pm$ 0.07 		  & $>$1.85 $\pm$ 0.07 	& 2.11 &  134 (82)	&108 &1516 	& 3.06E+15\\ 
FGS06  	& 08:44:56.6	&	+42:58:35.7	& 0.054     & 0.20 $\pm$ 0.12 		  & 2.11 $\pm$ 0.17  	   	& 0.72 & 21 (0)		&12	&330 	& 5.18E+13 \\ 
FGS07  	& 09:03:03.2	&	+27:39:29.4	& 0,489	&1.32 $\pm$ 0.33 		  & 1.96 $\pm$ 0.35  	   	& 1.68 & 28 (27)	&11	&926	& 8.94E+14 \\ 
FGS08* 	& 09:48:29.0	&	+49:55:06.7	& 0,409	&$>$2.12 $\pm$ 0.16 	  & $>$2.17 $\pm$ 0.14   	& 1.29 & -		&-	& - 		&  - \\ 
FGS09  	& 10:43:02.6	&	+00:54:18.3	& 0,125	&0.40 $\pm$ 0.30 		  & $>$0.68 $\pm$ 0.31  	& 1.55 & 58 (0)		& 11	&493	& 2.42E+14\\ 
FGS10** 	& 10:54:52.0	&	+55:21:12.5	& 0,468	&2.12 $\pm$ 0.33 		  & 2.24 $\pm$ 0.29 	  	& 1.43 &  116 (115)	& 78	&  969	& 8.32E+14 \\ 
FGS11 	&  11:14:39.8	&	+40:37:35.2	& 0,202	&$>$0.62 $\pm$ 0.11 	  & $>$1.03 $\pm$ 0.06  	& 1.34 & 47 (0)		& 0	& - 		&  - \\ 
FGS12 	& 11:21:55.3	&	+10:49:23.2	& 0,240	&1.61 $\pm$ 0.19		  & $>$2.00 $\pm$ 0.20 	& 1.34 &  54 (53)	& 24	&378 	&1.22E+14\\ 
FGS13 	& 11:41:28.3	&	+05:58:29.5	& 0,188	&1.23 $\pm$ 0.27		  & $>$1.80 $\pm$ 0.27 	& 1.19 &  40 (39)	& 14	&937 	&6.70E+14\\ 
FGS14* 	& 11:46:47.6	&	+09:52:28.2	& 0,221	&1.96 $\pm$ 0.29 		  & 2.43 $\pm$ 0.35 	   	& 1.45 &  78 (77)	& 40	& 774	& 5.55E+14 \\ 
FGS15$^a$ 	&11:48:03.8	&	+56:54:25.6	& 0,105	& 1.83 $\pm$ 0.09 		  & 2.27 $\pm$ 0.05  		& 0.98 &  63 (50) 	& 13	& -		& - \\ 
FGS16 	&11:49:15.0	&	+48:11:04.9	& 0,283	& $>$0.98 $\pm$ 0.33 	  & $>$1.13 $\pm$ 0.32  	& 1.45 & - 			&  -	& -		& - \\ 
FGS17* 	& 12:47:42.1	&	+41:31:37.7	& 0,155	&1.96 $\pm$ 0.55		  & $>$2.7 $\pm$ 0.23 	& 0.87 & - 			&  -	& -		& - \\ 
FGS18 	& 13:00:09.4	&	+44:43:01.3	& 0,233	&$>$1.41 $\pm$ 0.27 	  & $>$1.72 $\pm$ 0.29   	& 1.17 &- 			&  -	& - 		& -  \\ 
FGS19 	&13:35:60.0	&	-03:31:29.2	& 0,177	& 1.35 $\pm$ 0.23 		  & 1.97 $\pm$ 0.28 	   	& 1.37 &  57 (43)	& 25	& 978	& 8.35E+14\\ 
FGS20* 	& 14:10:04.2	&	+41:45:20.9	& 0,094	&2.17 $\pm$ 0.15 		  & $>$2.46 $\pm$ 0.14  	& 0.74 &  12 (0)		&10	& 578	& 1.63E+14\\ 
FGS21 	& 14:45:16.9	&	+00:39:34.3	& 0,306	&$<$0.00 $\pm$ 0.19 	  & $>$0.84 $\pm$ 0.26     & 1.53 & - 		&  -	& -		& - \\ 
FGS22 	& 14:53:59.0	&	+48:24:17.1	& 0,146	&1.49 $\pm$ 0.14 		  & 2.28 $\pm$ 0.14 	   	& 0.87 &  60 (57)	& 29	& 323 	&5.92E+13 \\ 
FGS23* 	& 15:29:46.3	&	+44:08:04.2	& 0,148	&1.87 $\pm$ 0.18 		  & $>$2.64 $\pm$ 0.14  	& 1.02 &  63 (60)	& 45	&659	& 2.86E+14 \\ 
FGS24 	& 15:33:44.1	&	+03:36:57.5	& 0,293	&0.33 $\pm$ 0.15 		  & 1.08 $\pm$ 0.20  	   	& 1.49 &  73 (69)	& 55	&780 	&5.75E+14 \\ 
FGS25 	& 15:39:50.8	&	+30:43:04.0	& 0,097	&1.12 $\pm$ 0.22 		  & 1.68 $\pm$ 0.29 	   	& 1.50 &  	70 (0)	& 25	&645 	&4.04E+14 \\ 
FGS26* 	& 15:48:55.9	&	+08:50:44.4	& 0,072	&1.18 $\pm$ 0.20 		  & $>$3.22 $\pm$ 0.19 	& 0.90 &  38 (0)		& 20	&675 	& 2.67E+14 \\ 
FGS27* 	& 16:14:31.1	&	+26:43:50.4	& 0,184	& 1.61 $\pm$ 0.22 	 	  & 2.64 $\pm$ 0.21  	   	& 1.26 &  	94 (88)	& 66	&910	& 6.69E+14 \\ 
FGS28*$^a$ 	& 16:37:20.5	&	+41:11:20.3	& 0,032	&$>$3.28 $\pm$ 0.07 	  & $>$3.68 $\pm$ 0.08   	& 0.47 &	76 (27)	& 1	& -		&  -\\ 
FGS29* 	& 16:47:02.1	&	+38:50:04.3	& 0,135	&1.81 $\pm$ 0.21 		  & 2.55 $\pm$ 0.22  	   	& 0.89 &  	46 (42)	& 18	&408	& 9.66E+13 \\ 
FGS30* 	& 17:18:11.9	&	+56:39:56.1	& 0,114	&1.84 $\pm$ 0.14 		  & 2.08 $\pm$ 0.14 	   	& 1.47 &  	63 (28)	& 39	&765	& 5.57E+14 \\ 
FGS31 	& 17:20:10.0	&	+26:37:32.1	& 0,159	&$>$1.04 $\pm$ 0.25 	  & $>$1.40 $\pm$ 0.23  	& 2.02 &  132 (89)	& 80	&1064	& 1.46E+15 \\ 
FGS32* 	& 17:28:52.2	&	+55:16:40.8	& 0,148	&$>$1.28 $\pm$ 0.26 	  & $>$2.48 $\pm$ 0.16  	& 0.78 &	-		& -	& - &  - \\ 
FGS33 	& 22:56:30.0	&	-00:32:10.8	& 0,224	&$>$1.11 $\pm$ 0.14 	  & $>$1.15 $\pm$ 0.13   	& 1.24 &	-		& -	& - &  - \\ 
FGS34* 	& 23:58:15.1	&	+15:05:43.6	& 0,178	&1.82  $\pm$ 0.28 		  & $>$ 3.09 $\pm$ 0.36    & 1.00 &  	52 (47)	& 22	&360 & 8.36E+13\\  
\hline

\end{tabularx}
\end{center}
\tiny Note: Col. (1): system number as in \citet{santos07}; Col (2): right ascension of the BGG; Col (3): declination of the BGG; Col (4): redshift of the BGG; Col (5): gap in magnitude between the two brightest member galaxies; Col (6): gap in magnitude between the brightest and the fourth brightest member galaxies; Col (7):estimation of the virial radius of the system derived from L$_X$; Col (8): number of velocities available within $R_{{\rm 200}}$. The number of new velocities from our own observations is given in parenthesis; Col (9): number of spectroscopically confirmed members; Col (10):  velocity dispersion of the system ; Col (11): mass of the system within 0.5 $R_{{\rm 200}}$, from Eq. \ref{eq:mass}.\\
\tiny * Fossil system according to our definition.
\\
\tiny ** For homogeneity, we recomputed the gaps for this system which we analised in \citet{aguerri11}. In particular, we used 0.5 $R_{200}$ and only three different magnitudes to calculate $\Delta m_{12}$ and $\Delta m_{14}$.
\\
\tiny $^a$ System with not reliable membership determination for which we did not compute $\sigma _v$ and $M_{{\rm v}}$.
\end{table*}

\subsection{L$_{\rm X}$-$\sigma_v$ relation}

\begin{figure}
\centering
\includegraphics[width=0.4\textwidth]{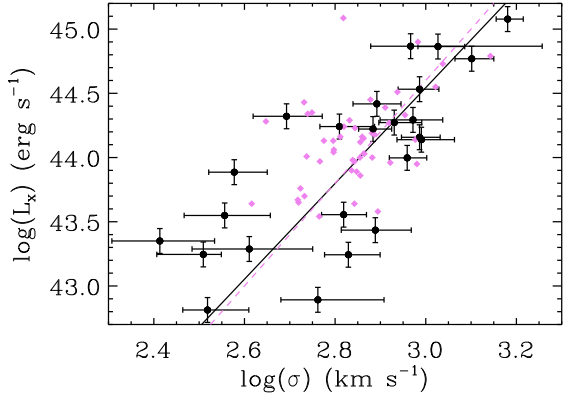}
\caption{L$_{\rm X}$-$\sigma _{{\rm v}}$ relation for the FGs candidates for which we were able to determine a velocity dispersion. The black circles represent our data and violet diamonds are taken from \citet{cava09} for comparison. The black solid line is our best fit and the violet dashed line is the best fit from \citet{cava09}.}
\label{lx-sigma}
\end{figure}

Once we obtained the luminosity in X-ray and velocity dispersion of the galaxies, we were able to evaluate the $L_{{\rm X}}-\sigma_{v}$ relation. This relation is connected with the formation of the cluster. In fact, theoretical predictions based on purely gravitational collapse lead to $L_{\rm X} \propto \sigma_{v}^4$. There are several observational studies of this relation, the majority of them finding values between 4 and 5.3 for the slope \citep[][]{quintana82,edge91,mul98,borgani99,xue00,mah01,Girardi_rad,ort04,hilton05}. In Fig. \ref{lx-sigma} we show the distribution and best-fit to our data. We found a slope consistent within the errors to the theoretical predictions. We derived the best-fit $L_{{\rm X}}-\sigma_{{\rm v}}$ relation using the FITEXY algorithm in IDL\footnote{Interactive Data Language is distributed by ITT Visual Information Solution. It is available from http://www.ittvis.com/} which account for measurements uncertainties in both variables.The ROSTAT package gave us the uncertainties in the velocity dispersion measurements, while for the X-ray luminosity we used the counts uncertainties listed by RASS-BCS/FSC and computed the relative error.  The same relative error was assumed for $L_{\rm X}$ and we found a median value of $\sigma_{L{\rm X}}\sim 25\%$.
Our best-fitting relation is:

\begin{equation}
{\rm log}(L_{{\rm X}})=(33.35\pm0.73) + (3.72\pm0.26) \,{\rm log}(\sigma_{{\rm v}})
\end{equation}
This relation is shown in Fig. \ref{lx-sigma} together with the $L_{\rm X}-\sigma_{{\rm v}}$ relation for the WINGS nearby cluster sample \citep{fasano06,cava09}. The two relations are in good agreement with one another. 

\subsection{Fossilness determination}
\label{fossilness}
\begin{figure}
\centering
\includegraphics[width=0.5\textwidth]{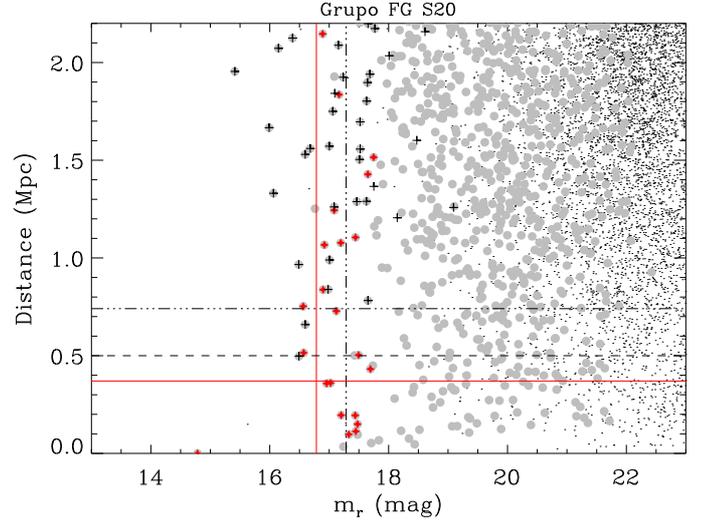}
\caption{Magnitude-distance diagram for the group FGS20. The distance from the BGG is given as a function of the magnitude for each galaxy. The black points represent all the galaxies within the FoV, grey circles are the target galaxies (see Sec. \ref{obs} for details), red stars represent the spectroscopically confirmed members, and green crosses are the spectroscopically confirmed non-members. The red solid horizontal line corresponds to 0.5 $R_{200}$, black dashed-dotted line marks $R_{200}$, and black dashed line represents 0.5 Mpc which is the limit used by \citet{santos07} to define the fossilness of the group. The red solid and dashed-dotted vertical lines indicate the 2-mag and 2.5-mag gaps from the BGG which determine the fossilness of the group following the criteria by \citet{jones03} and \citet{dariush10}, respectively.}
\label{mag-dist}
\end{figure}

A fossil system is defined as having a large gap in magnitude in the $r-$band between the two brightest members of the system, namely larger than 2 magnitues within 0.5 $R_{200}$. We calculated the distance from the BGG and magnitude for each galaxy to verify the fossilness criterium of our sample. In this way, we obtained a diagram (such as that shown in Fig. \ref{mag-dist}) that allowed us to constrain the main properties of the system, such as the magnitude gap, virial radius, cluster membership, and magnitude of the BGG. 

The magnitudes of the galaxies were obtained from SDSS-DR7 and our own photometry (see Sec. \ref{opt}). We used the extinction corrected Petrosian and model $r$-band magnitudes for all the objects in the SDSS-DR7 database. In addition, we have our own photometry for 30 out of 34 galaxy systems. Our photometry only covers the central regions of the clusters but it is about 2 magnitudes deeper in the $r-$band (see Fig. \ref{mag-dist}). We ran SExtractor \citep{bertin96} on our images in order to obtain the $r-$band  MAG-BEST\footnote{MAG-BEST is an aperture magnitude enclosing the total light of the galaxy. It usually coincides with MAG-AUTO, which is the best total magnitude provided for galaxies by SExtractor. The latter provides MAG-ISOCOR instead of MAG-AUTO if the galaxy is at least 10\% contaminated by another object.} magnitude.

Determining the magnitude of the BGGs is not straightforward. In SExtractor a successful deblending strongly depends on both the angular size of the BGG and the number of its close satellites. In order to circumvent this problem, we recomputed the magnitude of the BGGs using an {\it ad hoc} procedure on our images. In particular, we modeled the light of the BGGs by masking its close satellites. The model was done in IRAF by using the {\tt bmodel} task, which adopts as input the isophotal fit of the galaxy provided by {\tt ellipse} \citep[][]{ellipse}. The magnitude of the BGGs was computed using these uncontaminated models. Besides, the modeled light of the BGGs was subtracted from the original images and the magnitudes of all other galaxies were obtained by running SExtractor in the resulting images. 
SDSS photometry suffers from both deblending and overestimation of sky levels near bright galaxies \citep{blanton11}.  

The final magnitude of each galaxy was obtained by averaging the available magnitudes to have a more realistic estimate of the uncertainty. This was computed as the RMS of the available magnitudes. Table \ref{globprop} presents the value of the gap between the two brightest ($\Delta m_{12}$) and the first and fourth ranked galaxies ($\Delta m_{14}$) within 0.5 $R_{200}$ for each system. Some gaps are marked as lower limits because our spectroscopy failed to determine a redshift for some "bright" target galaxy (see Sec.\ref{compl} for details). For this reason we were not able to assign a membership to these objects. 

According to \citet{jones03} and \citet{dariush10}, a system if fossil if $\Delta m_{12} \ge 2.0$ or $\Delta m_{14} \ge 2.5$ mag, respectively. We considered as fossil systems those  that satisfy at least one of the previous criteria taking into account the errors in the magnitude gaps determination. More explicitly, a system is fossil if $\Delta m_{12}+ \epsilon_{12} \geq 2.0$ or $\Delta m_{14}+ \epsilon_{14} \geq 2.5$, being $\epsilon_{12}$ and $\epsilon_{14}$ the $1 \sigma$ uncertainties in the magnitude gaps. In Table 1 we highlighted the 15 systems that follow the previous criteria.

Notice that the two methods find 12 and 13 fossil systems respectively, despite \citet{dariush10} claimed that their method is expected to find 50\% more fossil systems than \citet{jones03}.

The fossil definitions take into account not only the magnitude gaps but also the virial radius of the system. Thus, uncertainties in the $R_{200}$ determination reflect in uncertainties in the fossil classification. Therefore, we computed the variation in the number of fossil systems taking into account  a 25\% uncertainty in $R_{{\rm 200}}$ in agreement with \citet{girardi14}. The number of fossil systems is  $15^{+8}_{-4}$. The upper limit gives the number of fossil systems for a 25\% smaller ${\rm R}_{200}$. Similarly, the lower limit corresponds to the number of fossil systems for a 25\% larger ${\rm R}_{200}$. 

\subsection{Correlations with the magnitude gaps}

\begin{figure}
\centering
\includegraphics[width=0.5\textwidth]{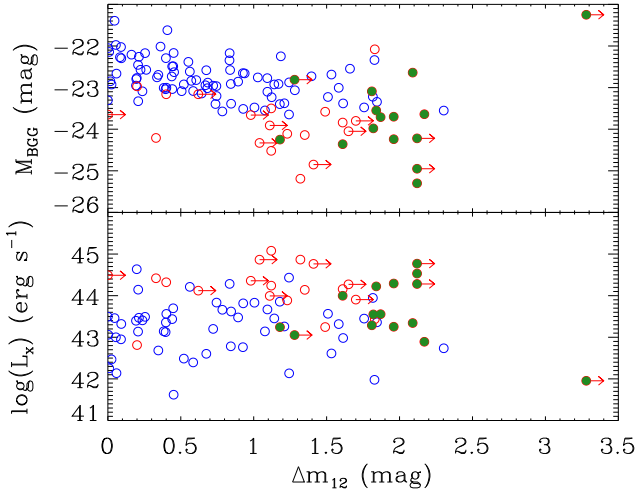}
\caption{Absolute $r$-band magnitude of the BGG (top panel) and X-ray luminosity of the system (bottom panel) as a function of the gap in magnitude between the two brightest galaxies of the systems studied in this paper (red open circles) and in  \citet[][blue open circles]{aguerri07}. The green filled circles are the genuine fossil systems. The points with a right arrow are those systems for which the magnitude gap is a lower limit.}
\label{plot_deltam}
\end{figure}

In Fig. \ref{plot_deltam} we show $\Delta m_{12}$  as a function of both the absolute $r-$band magnitude of the BGG (M$_{{\rm BGG}}$) and X-ray luminosity of the system. Our FG candidates show mainly $\Delta m_{12}>1$. In order to have more clusters in the range $0<\Delta m_{12}<1$, we included the sample of nearby ($z<0.1$) galaxy clusters from \citet{aguerri07}. Their $\Delta m_{12}$ was obtained from spectroscopically confirmed members within 0.5 $R_{{\rm200}}$ once we applied the same evolutionary and K corrections that we used for the \citet{santos07} sample. For both relations we computed the Spearman correlation coefficients. We found a strong correlation (significance $> 3\sigma$) between $\Delta m_{12}$ and M$_{{\rm BGG}}$. On average, the larger is $\Delta m_{12}$ the brightest is the central objects. The M$_{\rm BGG}$ and $\Delta m_{12}$ correlation is somewhat expected in the classical scenario of the formation of fossil systems, in which the central galaxy has grown due to merging of nearby $L^\ast$ galaxies. A similar correlation was also observed for central galaxies in other cluster samples \citep[see][and references therein]{ascaso11}. On the contrary, the relationship between log(L$_{\rm X}$) and $\Delta m_{12}$ is weaker (significance $< 2\sigma$). 

\begin{figure} 
\centering
\includegraphics[width=0.5\textwidth]{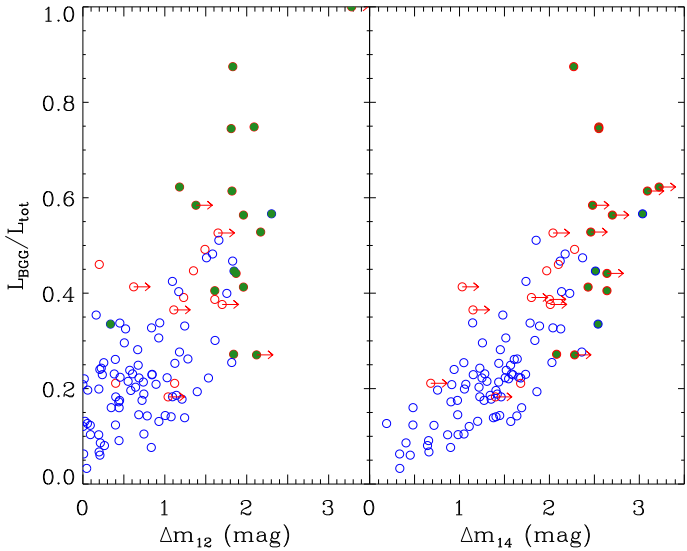}
\caption{Fraction of light of the BGG as function of $\Delta m_{12}$ and $\Delta m_{14}$ for the systems studied in this paper (red open circles) and in  \citet[][blue open circles]{aguerri07} with $z\leq0.25$. The green filled circles represent the genuine fossil systems. The points with a right arrow are those systems for which the magnitude gap is a lower limit.}
\label{frac_vs_deltam}
\end{figure}

We also analysed  the  fraction of total optical luminosity contained within the central galaxy ($L_{{\rm BGG}}/L_{{\rm tot}}$) as a function of the magnitude gaps $\Delta m_{12}$ and $\Delta m_{14}$. In this case, the total luminosity $L_{{\rm tot}}$ represents the sum of the luminosities of all the galaxies with $|(g-r)-(g-r)_{BGG}| \leq 0.2$, $M_{{\rm r}} \leq -20.0$, and  within 0.5 $R_{200}$. We limited this analysis to systems with $z\leq 0.25$ because for farther systems we were unable to reach $M_{\rm r}=-20.0$. Figure \ref{frac_vs_deltam} shows a clear correlation (Spearman test significance $> 3 \sigma$) between $L_{{\rm BGG}}/L_{{\rm tot}}$ and the two magnitude gaps. Fossil systems are, on average, those objects with a larger fraction of light in the BGG. Nevertheless, they are characterized by a large range of $L_{{\rm BGG}}/L_{{\rm tot}}$ values (0.25 $< L_{{\rm BGG}}/L_{{\rm tot}} <$ 0.75). Most of the systems with $L_{{\rm BGG}}/L_{{\rm tot}} >$ 0.5 are fossil ones. Similar relations were also found in other fossil samples \citep[see][]{harrison12} and recently \citet{shen13} suggest that the growth in mass of the BGGs is directly correlated with $\Delta m_{12}$ and that this correlation is necessary to justify the BGGs over luminosity.

Fossil systems represent always extreme cases in the correlations. However, Fig. \ref{plot_deltam} and Fig.\ref{frac_vs_deltam} indicate that not all the properties of the clusters depend on $\Delta m_{12}$. 

\subsection{Correlation with the velocity dispersion}

\begin{figure}
\centering
\includegraphics[width=0.5\textwidth]{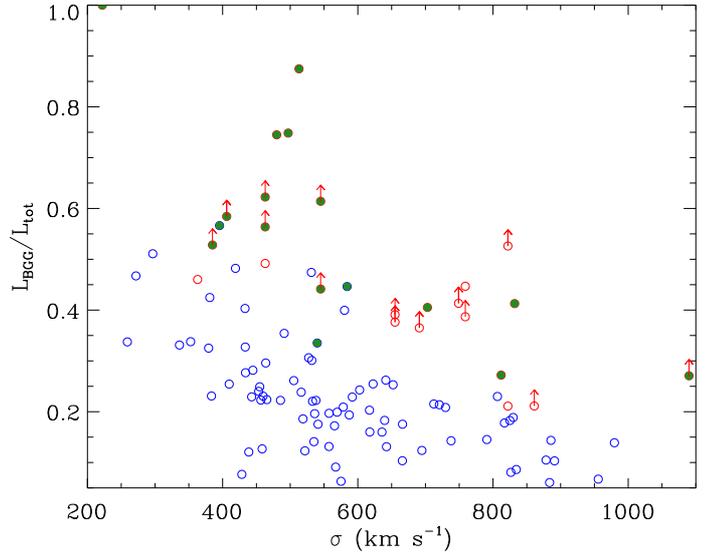}
\caption{Fraction of light of the BGG as function of the LOS velocity dispersion (mass) of the system. The symbols and colors are the same of Fig. \ref{frac_vs_deltam}. Upward arrows indicate those systems for which the magnitude gaps (and thus the L$_{{\rm BGG}}/$L$_{{\rm tot}}$) represent lower limits.}
\label{frac_vs_sigma}
\end{figure}

In Fig. \ref{frac_vs_sigma} we show the correlation between the fraction of light enclosed in the BGG ($L_{{\rm BGG}}/L_{{\rm tot}}$) as function of the LOS velocity dispersion $\sigma _{{\rm v}}$ of the cluster for the same sample of Fig. \ref{plot_deltam}. This is a well known correlation which was originally reported by \citet{linmohr04}. They argued that this correlation, together with the correlation between the luminosity of the BGGs and the mass of the system, indicates that BGGs grow by merging galaxies. In addition, they claimed that the decreasing of the BGG luminosity fraction with cluster mass indicates that the rate of luminosity growth of the BGGs is slower than the rate at which clusters acquire galaxies from the field.

Fig. \ref{frac_vs_sigma} clearly shows that fossil systems delineate the upper envelope of the expected trend of the $L_{{\rm BGG}}/L_{{\rm tot}}-\sigma_{{\rm v}}$ relation of non-fossil systems. So, for a given velocity dispersion (or mass), fossil systems have a larger fraction of light enclosed in the BGG. Following Lin \& Mohr, we infer that the growth rate of BGGs in fossil systems is larger than that of BGGs in non-fossil systems. Notice that non-fossil systems which are located in the upper envelope of Fig. \ref{frac_vs_sigma} have either large gaps in magnitude or gaps calculated as lower limits only. The former are systems dominated by the BGG which are classified as non-fossil systems only due to the arbitrariness of the fossilness criteria. The latter are expected to be genuine fossil systems, but  for which further investigation is needed to constrain $\Delta m_{12}$ and $\Delta m_{14}$.  

\section{Discussion}
\label{disc}

\subsection{Differences with \citet{santos07}}
\label{diff}
We carefully analysed the sample of FG candidates presented by \citet{santos07}. As they claimed, the number of fossil systems depends critically on the parameters adopted for measuring the magnitude gaps. Hence, it is very important to highlight the differences between their and our methodology. First of all, \citet{santos07} used a fixed radius to look for the second brightest galaxies, namely 0.5 Mpc, whereas we were able to compute the virial radius for each cluster from its X-ray luminosity. The value of 0.5 $R_{200}$ that we obtained for the \citet{santos07} candidates varies between 0.29  Mpc and 1.02 Mpc, with a mean value of 0.63 Mpc. This means that, on average, the radius within which we looked for the second brightest galaxies is larger than the one adopted in \citet{santos07}. Therefore, we expect that not all the candidates proposed by \citet{santos07} are genuine fossil systems. 

Moreover, the procedure to define cluster membership adopted by \citet{santos07} suffers from three main problems. First, a small number of spectroscopic redshifts were available in the SDSS-DR5 for the 34 FG candidates. Second, \citet{santos07} considered as members all the galaxies with the spectroscopic redshift in the range $z_{\rm c} \pm \Delta z$, being $z_{\rm c}$ the cluster redshift and $\Delta z=0.002$. It is worth noting that $\Delta z$ was fixed for all the clusters, and did not take into account the differences in velocity dispersion (or mass) between the systems. Third, \citet{santos07} considered as cluster members also the galaxies without the spectroscopic redshift but with a photometric redshift in the range $z_{\rm c} \pm 0.035$. This photometric redshift window is too small to deal with the typical errors of 0.1 expected for SDSS-DR5 photometric redshifts. In addition, they only took into account galaxies with errors smaller than 0.1 in the photometric redshift.

On the contrary, we were able to obtain a number of spectroscopic velocities good enough to compute the LOS velocity dispersion of the systems and to accurately define the cluster membership. In particular, for 25 out of 34 systems the cluster membership was obtained using a two-step method of member selection which works both in the redshift space and projected space phase. This method is much more robust than the simple $z$-cut proposed by \citet{santos07}. Moreover, our photometric redshift criterion for the membership adopts a much larger window of $z_c \pm 0.15$. For all these reasons, it is not surprising that only half of the systems proposed by \citet{santos07} as FG candidates turned out to be genuine fossil systems according to our criteria. Finally, it is important to notice that there are 12 systems in the sample that are not fossils but for which the magnitude gaps we calculated are lower limits only. This means that, in principle, there could be other genuine fossil systems in the sample, which could be identified by completing the spectroscopic survey.

\subsection{Formation scenarios for fossil systems}
There are two models that are mainly used in the literature to explain the formation scenarios for fossil systems. The first and widely accepted one \citep[][among others]{donghia05,sommer06,vonbenda08,dariush10} is that fossil systems were assembled at a higher redshift than regular clusters and, due to particularly favorable conditions, they had enough time to merge $L^\ast$ galaxies in a single giant BGG. This process results in the observed gap in magnitude which is a consequence of the system evolution. The second scenario \citep[][]{mul99} suggests that the BGG was created in a single monolithic collapse and that the gap in magnitude is present from the beginning. The two formation processes are called {\it merging} and {\it failed group} scenario, respectively. 

\noindent
The observational properties of our systems are the following:

\begin{enumerate}[(a)]
\item Fossil systems are observed at any LOS velocity dispersion (mass) of the host DM halo (Table \ref{globprop}). Indeed, we have both fossil groups and fossil clusters.
\item Systems with larger $\Delta m_{12}$ have brighter BGGs (Fig. \ref{plot_deltam}, top panel).
\item Systems with larger $\Delta m_{12}$ or $\Delta m_{14}$ have a larger fraction of total galaxy light contained within the BGG (Fig. \ref{frac_vs_deltam}).
\item Fossil systems follow the same $L_{{\rm BGG}}/L_{{\rm tot}}-\sigma_{{\rm v}}$ relation of non-fossil systems. However, fossil systems are extreme cases of this relation and show larger $L_{{\rm BGG}}/L_{{\rm tot}}$ for a fixed LOS velocity dispersion (Fig. \ref{frac_vs_sigma}).
\item Fossil systems lie on the same $L_{{\rm X}}-\sigma _{{\rm v}}$ relation of non-fossil systems.
\end{enumerate}

All these facts can be interpreted in the terms of the merging scenario: the fossil systems are the result of massive merger episodes in the early Universe due to the fact that their galaxies follow more low angular momentum orbits than galaxies in non-fossil systems. In this case, the observed magnitude gap is an indication of the evolutionary state of the system. Thus, the systems with larger magnitude gaps are expected to have brighter central galaxies and larger fraction of light enclosed in their BGGs. The growth of the BGG in fossil systems is larger than in non-fossil ones, but it is expected to follow the same rules of normal clusters \citep[see][]{linmohr04}.  

\noindent
The properties we observed in fossil systems can also be explained by the failed group scenario.  
Nevertheless, there are other properties that seem to disfavor this scenario. For example, the monolithic formation of elliptical galaxies predicts strong metallicity gradients whereas the stellar population gradients are erased by mergers. In this context, \citet{paul13} found flat age and metallicity gradients for a sample of central galaxies in fossil systems,which are not compatible with the failed group scenario. Moreover, the bend observed at high masses in the scaling relations of early-type galaxies also suggests that fossil systems were formed by mergers \citep[see][]{bernardi11}. In particular, for the BGGs of our 34 FG candidates a two-phase merger scenario was proposed to explain their scaling relations. Indeed, \citep[][]{jairo12} claimed that these objects went through dissipational mergers in an early stage of their evolution and assembled the bulk of their mass through subsequent dry mergers. This process seems similar to the one proposed by \citet{diaz08}, in which the BGGs in fossil systems have undergo their last major merger later than in non-fossil systems.

\subsection{Transitional fossil systems}

The correlation between the gap in magnitude and absolute magnitude of the BGG suggests that the scenario that suggests the existence of transitional fossil systems \citep{vonbenda08} can not be applied to those systems with the brightest BGGs. In fact, the probability that two systems with such a bright central galaxy would merge is negligible. For this reason, we expect that current fossil systems hosting the most luminous BGGs will be fossil forever. Nevertheless, the \citet{vonbenda08} scenario could explain the case of FGS06. The BGG has a magnitude of $M_r=-22.88$ whereas the second ranked, located at $\sim 0.4$ $R_{200}$, has $M_r=-22.68$. These magnitude values are typical of central galaxies in groups and clusters (see Fig. \ref{plot_deltam}). The third, fourth and fifth ranked galaxies have magnitudes of $M_r=-21.19$, $-20.77$, and $-20.18$, respectively. Moreover, FGS06 is the only non-fossil system in our sample with an abrupt change in its magnitude gap if the second ranked galaxy would not be taken into account. In fact, the median value of the magnitude gap change for non-fossil systems in our sample is $0.2\pm 0.2$ mag, whereas FGS06 would suffer a 1.6 mag change. For these reasons, we suggest that FGS06 could be a good candidate for a transitional fossil systems as those described by \citet{vonbenda08}.


\section{Conclusions}

We characterized the sample of 34 FG candidates proposed by \citet{santos07} by using a unique collection of new optical photometric and spectroscopic data. This dataset was completed with SDSS-DR7 archival data. This large collection of radial velocities provided us robust cluster membership and global cluster properties for a subsample of 25 systems which were not available before. 

The fossilness determination of the 34 FG candidates was revisited. In particular, the magnitudes of the galaxies in each system were obtained by averaging three different magnitudes: Petrosian and model magnitudes from  SDSS-DR7 and MAG-BEST SExtractor magnitude from our data. This was done because the magnitude of the BGGs can be affected by close satellites both in the SDSS and SExtractor analyses. 
Therefore, we computed new magnitude gaps ($\Delta m_{12}$ and  $\Delta m_{14}$) within 0.5 $R_{200}$ for each system. The systems with $\Delta m_{12} \ge 2$ or $\Delta m_{14} \ge 2.5$ mag within the errors were classified as fossils. By applying this criterion, the total number of fossil systems in the sample is $15^{+8}_{-4}$. The uncertainties in the total number of fossil systems reflect the uncertainties in the $R_{200}$ determination. Moreover, there are 12 systems for which one or both the magnitude gaps are lower limits. For these systems, a more extended spectroscopic survey is needed in order to define their fossilness.

We derived the main observational properties of the fossil systems in our sample. The fossil systems span a wide range of masses and we can confirm the existence of genuine fossil clusters in our sample. In particular, five fossil systems have LOS velocity dispersions $\sigma _{{\rm v}} > 700$ km s$^{-1}$, from both the $L_X$ luminosity and "shifting gapper" procedure. 
Clear correlations were found between the magnitude gaps and luminosity of the BGGs. In particular, the systems with larger $\Delta m_{12}$ have brighter BGGs, and the systems with larger $\Delta m_{12}$ or $\Delta m_{14}$ have larger fraction of the total galaxy light in the BGGs. The fossil systems also follow the same $L_{{\rm BGG}}/L_{{\rm tot}}-\sigma_{{\rm v}}$ relation of non-fossil systems. Nevertheless, they are extreme cases in the studied relations. In particular, the fossil systems have brighter BGGs than normal systems for any given LOS velocity dispersion (mass).

All these properties can be explained by the two mainly accepted proposed scenarios of formation of fossil structures and thus are not conclusive in this sense. Nevertheless, we suggest that fossil systems with very bright central galaxies are not transitional phases of regular clusters and groups because, if this was the case, we should find systems with small gaps but very bright and massive central galaxies. These systems are not observed because the probability that two systems with such a bright BGG would merge is negligible. On the contrary, the systems with fainter BGGs possibly experienced a transitional fossil stage, which ended with the merging of another galaxy system. This could be the case of FGS06.

The FOGO project will continue in the next future by analyzing other observational properties of fossil systems. In a forthcoming paper we will focus on the LFs of fossil and normal systems. This analysis will be crucial for the understanding of the formation and evolution of these galaxy aggregations, because the LFs of fossil systems in the merging scenario are expected to have a lack of $L^\ast$ galaxies. In contrast, the failed group formation scenario expects to find differences between fossil and normal systems in both the bright and faint ends of the LFs.

\begin{acknowledgements}
We would like to thank the anonymous referee for the useful comments which helped us to improve the paper. This work was partially funded by the Spanish MICINN (grant AYA2010-21887-C04-04), and the local Canarian Government (grant ProID20100140). This article is based on observations made with the Isaac Newtown Telescope, Nordic Optical Telescope, and Telescope Nazionale Galileo operated on the island of La Palma by the Isaac Newton Group, the Nordic Optical Telescope Scientific Association and the Fundación Galileo Galilei of the INAF (Istituto Nazionale di Astrofisica) respectively, in the Spanish Observatorio del Roque de los Muchachos of the Instituto de Astrofísica de Canarias. E.D. gratefully acknowledges the support of the Alfred P. Sloan Foundation. M.G. acknowledges financial support from the MIUR PRIN/2010-2011 (J91J12000450001). E.M.C. is supported by Padua University (grants 60A02-1283/10,5052/11, 4807/12). JIP and JVM acknowledge financial support from the Spanish MINECO under grant AYA2010-21887-C04-01, and from Junta de Andalucía Excellence Project PEX2011-FQM7058. JMA acknowledges support from the European Research Council Starting Grant (SEDmorph; P.I. V. Wild)
\end{acknowledgements}


\begin{thebibliography}{}
\bibitem[Abazajian et al.(2009)]{DR7} Abazajian, K.~N., 
Adelman-McCarthy, J.~K., Ag{\"u}eros, M.~A., et al.\ 2009, \apjs, 182, 543 
\bibitem[Adelman-McCarthy et al.(2007)]{DR5} 
Adelman-McCarthy, J.~K., Ag{\"u}eros, M.~A., Allam, S.~S., et al.\ 2007, 
\apjs, 172, 634 
\bibitem[Adelman-McCarthy et al.(2008)]{DR6} 
Adelman-McCarthy, J.~K., Ag{\"u}eros, M.~A., Allam, S.~S., et al.\ 2008, 
\apjs, 175, 297 
\bibitem[Aguerri et 
al.(2007)]{aguerri07} Aguerri, J.~A.~L., S{\'a}nchez-Janssen, R., \& Mu{\~n}oz-Tu{\~n}{\'o}n, C.\ 2007, \aap, 471, 17
\bibitem[Aguerri et 
al.(2011)]{aguerri11} Aguerri, J.~A.~L., Girardi, M., Boschin, W., et al.\ 2011, \aap, 527, A143 
\bibitem[Arnaud et 
al.(2005)]{Arnaud05} Arnaud, M., Pointecouteau, E., \& Pratt, G.~W.\ 2005, \aap, 441, 893 
\bibitem[Ascaso et al.(2011)]{ascaso11} Ascaso, B., Aguerri, 
J.~A.~L., Varela, J., et al.\ 2011, \apj, 726, 69
\bibitem[Bardelli et al.(1994)]{bar94} Bardelli, S., Zucca, 
E., Vettolani, G., et al.\ 1994, \mnras, 267, 665 
\bibitem[Barrena et 
al.(2009)]{bar09} Barrena, R., Girardi, M., Boschin, W., \& Das{\'{\i}}, M.\ 2009, \aap, 503, 357 
\bibitem[Beers et al.(1990)]{beers90} Beers, T.~C., Flynn, K., 
\& Gebhardt, K.\ 1990, \aj, 100, 32
\bibitem[Bernardi et al.(2011)]{bernardi11} Bernardi, M., Roche, 
N., Shankar, F., \& Sheth, R.~K.\ 2011, \mnras, 412, 684
\bibitem[Bertin 
\& Arnouts(1996)]{bertin96} Bertin, E., \& Arnouts, S.\ 1996, \aaps, 117, 393
\bibitem[Blanton et al.(2011)]{blanton11} Blanton, M.~R., Kazin, 
E., Muna, D., Weaver, B.~A., \& Price-Whelan, A.\ 2011, \aj, 142, 31
\bibitem[B{\"o}hringer et al.(2000)]{Bor00} B{\"o}hringer, 
H., Voges, W., Huchra, J.~P., et al.\ 2000, \apjs, 129, 435
\bibitem[B{\"o}hringer et 
al.(2004)]{Bor04} B{\"o}hringer, H., Schuecker, P., Guzzo, L., et al.\ 2004, \aap, 425, 367 
\bibitem[B{\"o}hringer et 
al.(2007)]{Bor07} B{\"o}hringer, H., Schuecker, P., Pratt, G.~W., et al.\ 2007, \aap, 469, 363
\bibitem[Borgani et al.(1999)]{borgani99} Borgani, S., Girardi, 
M., Carlberg, R.~G., Yee, H.~K.~C., \& Ellingson, E.\ 1999, \apj, 527, 561 
\bibitem[Boylan-Kolchin et al.(2008)]{boylan08} Boylan-Kolchin, 
M., Ma, C.-P., \& Quataert, E.\ 2008, \mnras, 383, 93
\bibitem[Carlberg et al.(1997)]{carlberg97} Carlberg, R.~G., Yee, 
H.~K.~C., \& Ellingson, E.\ 1997, \apj, 478, 462 
\bibitem[Cava et al.(2009)]{cava09} Cava, A., Bettoni, D., Poggianti, B.~M., et al.\ 2009, \aap, 495, 707 
\bibitem[Coziol et al.(2009)]{coziol09} Coziol, R., Andernach, 
H., Caretta, C.~A., Alamo-Mart{\'{\i}}nez, K.~A., 
\& Tago, E.\ 2009, \aj, 137, 4795
\bibitem[Cypriano et al.(2006)]{cypriano06} Cypriano, E.~S., 
Mendes de Oliveira, C.~L., \& Sodr{\'e}, L., Jr.\ 2006, \aj, 132, 514 
\bibitem[Dahle et al.(2002)]{dahle} Dahle, H., Kaiser, N., 
Irgens, R.~J., Lilje, P.~B., \& Maddox, S.~J.\ 2002, \apjs, 139, 313
\bibitem[Danese et 
al.(1980)]{danese80} Danese, L., de Zotti, G., \& di Tullio, G.\ 1980, \aap, 82, 322 
\bibitem[Dariush et al.(2007)]{dariush07} Dariush, A., 
Khosroshahi, H.~G., Ponman, T.~J., et al.\ 2007, \mnras, 382, 433
\bibitem[Dariush et al.(2010)]{dariush10} Dariush, A.~A., 
Raychaudhury, S., Ponman, T.~J., et al.\ 2010, \mnras, 405, 1873
\bibitem[D{\'{\i}}az-Gim{\'e}nez et 
al.(2008)]{diaz08} D{\'{\i}}az-Gim{\'e}nez, E., Muriel, H., \& Mendes de Oliveira, C.\ 2008, \aap, 490, 965 
\bibitem[D{\'e}mocl{\`e}s et 
al.(2010)]{democles10} D{\'e}mocl{\`e}s, J., Pratt, G.~W., Pierini, D., et al.\ 2010, \aap, 517, A52 
\bibitem[D'Onghia 
\& Lake(2004)]{donghia04} D'Onghia, E., \& Lake, G.\ 2004, \apj, 612, 628 
 \bibitem[D'Onghia et al.(2005)]{donghia05} D'Onghia, E., 
Sommer-Larsen, J., Romeo, A.~D., et al.\ 2005, \apjl, 630, L109
\bibitem[Edge 
\& Stewart(1991)]{edge91} Edge, A.~C., \& Stewart, G.~C.\ 1991, \mnras, 252, 414 
\bibitem[Eigenthaler 
\& Zeilinger(2012)]{paul12} Eigenthaler, P., \& Zeilinger, W.~W.\ 2012, \aap, 540, A134
\bibitem[Eisenstein et al.(2001)]{LRG} Eisenstein, D.~J., 
\bibitem[Eigenthaler 
\& Zeilinger(2013)]{paul13} Eigenthaler, P., \& Zeilinger, W.~W.\ 2013, \aap, 553, A99 
Annis, J., Gunn, J.~E., et al.\ 2001, \aj, 122, 2267 
\bibitem[Ellingson 
\& Yee(1994)]{ell94} Ellingson, E., \& Yee, H.~K.~C.\ 1994, \apjs, 92, 33 
\bibitem[Fadda et al.(1996)]{fadda96} Fadda, D., Girardi, M., 
Giuricin, G., Mardirossian, F., \& Mezzetti, M.\ 1996, \apj, 473, 670 
\bibitem[Falco et al.(1999)]{falco99} Falco, E.~E., Kurtz, 
M.~J., Geller, M.~J., et al.\ 1999, \pasp, 111, 438 
\bibitem[Fasano et 
al.(2006)]{fasano06} Fasano, G., Marmo, C., Varela, J., et al.\ 2006, \aap, 445, 805 
\bibitem[Girardi et al.(1996)]{girardi96} Girardi, M., Fadda, D., 
Giuricin, G., et al.\ 1996, \apj, 457, 61
\bibitem[Girardi et al.(1998)]{girardi98} Girardi, M., Giuricin, 
G., Mardirossian, F., Mezzetti, M., \& Boschin, W.\ 1998, \apj, 505, 74
\bibitem[Girardi 
\& Mezzetti(2001)]{Girardi_rad} Girardi, M., \& Mezzetti, M.\ 2001, \apj, 548, 79 
\bibitem[Girardi et al.(2006)]{A697} Girardi, M., Boschin, W., \& Barrena, R.\ 2006, \aap, 455, 45 
\bibitem[Girardi et al. (2014), in preparation]{girardi14}
\bibitem[Harrison et al.(2012)]{harrison12} Harrison, C.~D., 
Miller, C.~J., Richards, J.~W., et al.\ 2012, \apj, 752, 12 
\bibitem[Hilton et al.(2005)]{hilton05} Hilton, M., Collins, C., 
De Propris, R., et al.\ 2005, \mnras, 363, 661 
\bibitem[Huang et al.(2013a)]{huang13a} Huang, S., Ho, L.~C., 
Peng, C.~Y., Li, Z.-Y., \& Barth, A.~J.\ 2013, \apj, 766, 47 
\bibitem[Huang et al.(2013b)]{huang13b} Huang, S., Ho, L.~C., 
Peng, C.~Y., Li, Z.-Y., \& Barth, A.~J.\ 2013, \apjl, 768, L28 
\bibitem[Jedrzejewski(1987)]{ellipse} Jedrzejewski, R.~I.\ 
1987, \mnras, 226, 747
\bibitem[Jones et al.(2003)]{jones03} Jones, L.~R., Ponman, 
T.~J., Horton, A., et al.\ 2003, \mnras, 343, 627 
\bibitem[Kennicutt(1992)]{ken92} Kennicutt, R.~C., Jr.\ 1992, 
\apjs, 79, 255 
\bibitem[Khosroshahi et al.(2004)]{khos04} Khosroshahi, H.~G., 
Jones, L.~R., \& Ponman, T.~J.\ 2004, \mnras, 349, 1240
\bibitem[Khosroshahi et al.(2006)]{khos06} Khosroshahi, H.~G., 
Maughan, B.~J., Ponman, T.~J., \& Jones, L.~R.\ 2006, \mnras, 369, 1211 
\bibitem[Khosroshahi et al.(2007)]{khos07} Khosroshahi, H.~G., 
Ponman, T.~J., \& Jones, L.~R.\ 2007, \mnras, 377, 595
\bibitem[Klypin et al.(1999)]{klypin99} Klypin, A., Kravtsov, 
A.~V., Valenzuela, O., \& Prada, F.\ 1999, \apj, 522, 82 
\bibitem[La Barbera et al.(2009)]{labarbera09} La Barbera, F., de 
Carvalho, R.~R., de la Rosa, I.~G., et al.\ 2009, \aj, 137, 3942 
\bibitem[Lidman et al.(2013)]{lidman13} Lidman, C., Iacobuta, 
G., Bauer, A.~E., et al.\ 2013, \mnras, 433, 825
\bibitem[Lieder et 
al.(2013)]{lieder13} Lieder, S., Mieske, S., S{\'a}nchez-Janssen, R., et al.\ 2013, \aap, 559, A76
\bibitem[Lin 
\& Mohr(2004)]{linmohr04} Lin, Y.-T., \& Mohr, J.~J.\ 2004, \apj, 617, 879 
\bibitem[Malumuth et al.(1992) ]{mal92} Malumuth, E.~M., 
Kriss, G.~A., Dixon, W.~V.~D., Ferguson, H.~C., 
\& Ritchie, C.\ 1992, \aj, 104, 495
\bibitem[Mahdavi 
\& Geller(2001)]{mah01} Mahdavi, A., \& Geller, M.~J.\ 2001, \apjl, 554, L129 
\bibitem[Markevitch(1998)]{mark98} Markevitch, M.\ 1998, \apj, 
504, 27
\bibitem[Marrone et al.(2012)]{marrone12} Marrone, D.~P., Smith, 
G.~P., Okabe, N., et al.\ 2012, \apj, 754, 119
\bibitem[Materne(1979)]{Materne} Materne, J.\ 1979, \aap, 74, 235 
\bibitem[Merch{\'a}n 
\& Zandivarez(2005)]{merch05} Merch{\'a}n, M.~E., \& Zandivarez, A.\ 2005, \apj, 630, 759
\bibitem[Mendes de Oliveira et al.(2006)]{mendes06} Mendes de 
Oliveira, C.~L., Cypriano, E.~S., 
\& Sodr{\'e}, L., Jr.\ 2006, \aj, 131, 158 
\bibitem[Mendes de Oliveira et al.(2009)]{mendes09} Mendes de 
Oliveira, C.~L., Cypriano, E.~S., Dupke, R.~A., 
\& Sodr{\'e}, L., Jr.\ 2009, \aj, 138, 502 
\bibitem[M{\'e}ndez-Abreu et 
al.(2012)]{jairo12} M{\'e}ndez-Abreu, J., Aguerri, J.~A.~L., Barrena, R., et al.\ 2012, \aap, 537, A25\bibitem[Mukai(1993)]{muk93} Mukai, K.\ 1993, Legacy, vol.~3, 
p.21-31, 3, 21
\bibitem[Mulchaey 
\& Zabludoff(1998)]{mul98} Mulchaey, J.~S., \& Zabludoff, A.~I.\ 1998, \apj, 496, 73 
\bibitem[Mulchaey 
\& Zabludoff(1999)]{mul99} Mulchaey, J.~S., \& Zabludoff, A.~I.\ 1999, \apj, 514, 133 
\bibitem[Ortiz-Gil et al.(2004)]{ort04} Ortiz-Gil, A., Guzzo, 
L., Schuecker, P., B{\"o}hringer, H., 
\& Collins, C.~A.\ 2004, \mnras, 348, 325 
\bibitem[Osterbrock et al.(1996)]{osterbrock96} Osterbrock, D.~E., 
Fulbright, J.~P., Martel, A.~R., et al.\ 1996, \pasp, 108, 277 
\bibitem[Paolillo et al.(2001)]{paolillo01} Paolillo, M., Andreon, S., Longo, G., et al.\ 2001, \aap, 367, 59 
\bibitem[Pisani(1993)]{dedica2} Pisani, A.\ 1993, \mnras, 265, 
706
\bibitem[Pisani(1996)]{dedica1} Pisani, A.\ 1996, \mnras, 278, 
697
\bibitem[Ponman et al.(1994)]{ponman94} Ponman, T.~J., Allan, 
D.~J., Jones, L.~R., et al.\ 1994, \nat, 369, 462 
\bibitem[Popesso et 
al.(2005)]{pop05} Popesso, P., B{\"o}hringer, H., Romaniello, M., \& Voges, W.\ 2005, \aap, 433, 415 
\bibitem[Proctor et al.(2011)]{proctor11} Proctor, R.~N., de 
Oliveira, C.~M., Dupke, R., et al.\ 2011, \mnras, 418, 2054
\bibitem[Quintana 
\& Melnick(1982)]{quintana82} Quintana, H., \& Melnick, J.\ 1982, \aj, 87, 972 
\bibitem[Quintana et al.(2000)]{qui00} Quintana, H., 
Carrasco, E.~R., \& Reisenegger, A.\ 2000, \aj, 120, 511 
\bibitem[Santos et al.(2007)]{santos07} Santos, W.~A., Mendes de 
Oliveira, C., \& Sodr{\'e}, L., Jr.\ 2007, \aj, 134, 1551
\bibitem[Shen et al. (2013), in preparation]{shen13}
\bibitem[Sommer-Larsen(2006)]{sommer06} Sommer-Larsen, J.\ 2006, 
\mnras, 369, 958
\bibitem[Sun et al.(2004)]{sun04} Sun, M., Forman, W., 
Vikhlinin, A., et al.\ 2004, \apj, 612, 805 
\bibitem[Voevodkin et al.(2010)]{voev10} Voevodkin, A., 
Borozdin, K., Heitmann, K., et al.\ 2010, \apj, 708, 1376 
\bibitem[Voges et 
al.(1999)]{vog99} Voges, W., Aschenbach, B., Boller, T., et al.\ 1999, \aap, 349, 389 
\bibitem[Voges et al.(2000)]{vog00} Voges, W., Aschenbach, 
B., Boller, T., et al.\ 2000, \iaucirc, 7432, 1
\bibitem[The 
\& White(1986)]{pressure} The, L.~S., \& White, S.~D.~M.\ 1986, \aj, 92, 1248
\bibitem[Tonry 
\& Davis(1979)]{ton79} Tonry, J., \& Davis, M.\ 1979, \aj, 84, 1511
\bibitem[Voges et al.(1999)]{RASS} Voges, W., Aschenbach, B., Boller, T., et al.\ 1999, \aap, 349, 389 
\bibitem[von Benda-Beckmann et al.(2008)]{vonbenda08} von 
Benda-Beckmann, A.~M., D'Onghia, E., Gottl{\"o}ber, S., et al.\ 2008, 
\mnras, 386, 2345 
\bibitem[Xue 
\& Wu(2000)]{xue00} Xue, Y.-J., \& Wu, X.-P.\ 2000, \apj, 538, 65 
\bibitem[Zibetti et al.(2009)]{zibetti09} Zibetti, S., Pierini, 
D., \& Pratt, G.~W.\ 2009, \mnras, 392, 525 
 
\end{thebibliography}
\end{document}